\documentclass[a4paper]{article}
\usepackage[T1]{fontenc}
\usepackage[utf8]{inputenc}
\usepackage[italian,english]{babel}
\usepackage{amsmath}
\usepackage{graphicx}
\usepackage{anysize}
\usepackage{float}
\usepackage{amsmath}
\usepackage{amssymb}
\usepackage{amsfonts}
\usepackage{bm}
\usepackage{array}
\usepackage{graphicx}
\usepackage{epsfig}
\usepackage{multirow}
\usepackage{multicol}
\usepackage{longtable}
\usepackage{ifthen}
\usepackage{lscape}
\usepackage{alphalph}
\usepackage{attachfile2}
\usepackage {textcomp}
\graphicspath{{figs/}}
\normalmarginpar
\begin{document}
  \begin{titlepage}
  \begin{center}

\textbf{On the invariant quantities in an entropy balance and their impact on the stability of some physical systems}
\

A. Di Vita \footnote{Univ. Genova D.I.C.C.A. Genova, Italy}, 

  \end{center}

\begin{abstract}
Recently, it has been shown [M. Polettini et al., 12th Joint European Thermodynamics Conference, Brescia, Italy, July 1-5, 2013] that diffeomorphisms in the state space of a wide class of physical systems leave the amount $ \Pi $ of entropy produced per unit time inside the bulk of the system unaffected. Starting from this invariance, we show that if the boundary conditions allow the system to relax towards some final ('relaxed') state, then the necessary condition for the stability of the relaxed state against slowly evolving perturbations is the same for all the systems of this class, regardless of the detailed dynamics of the system, the amplitude of fluctuations around mean values, and the possible occurrence of periodic oscillations in the relaxed state. We invoke also no Onsager symmetry, no detailed model of heat transport and production, and no approximation of local thermodynamic equilibrium. This necessary condition is the constrained minimization of suitably time- and path-ensemble-averaged  $ \Pi $ [G. E. Crooks, Phys. Rev. \textbf{E} \textbf{61}, 3, 2361 (2000)], the constraints being provided by the equations of motion. Even if the latter may differ, the minimum property of stable relaxed states remains the same in different systems. Relaxed states satisfy the equations of motion; stable relaxed states satisfies also the minimum condition. (Thermodynamic equilibrium corresponds to $ \Pi \equiv 0 $ and satisfies trivially this condition). Our class of physical systems includes both some mesoscopic systems described by a probability distribution function which obeys a simple Fokker-Planck equation and some macroscopic, classical fluids with no net mass source and no mass flow across the boundaries; the problems of stability of these systems are dual to each other. As particular cases, we retrieve some necessary criteria for stability of relaxed states in both macroscopic and mesoscopic systems. Retrieved criteria include (among others) constrained minimization 
of dissipated power for plasma filaments in a Dense Plasma Focus [A. Di Vita, Eur. Phys. J. D \textbf{54}, 451 (2009)], of electrical resistivity in plasmoids [L. Comisso et al., Phys. Plasmas \textbf{23}, 100702 (2016)], of time- and path-ensemble-averaged dissipated power in flashing ratchets [J. Dolbeault et al., Technical Report 0244, CEREMADE (2002)] and of the so-called 'effective temperature' in Brownian motors [A. Feigel et al., arXiv:1312.5279v1 cond-mat.stat-mech 18 Dec 2013], as well as the scaling of the average magnetic field to the reciprocal of Rossby number in dynamo-affected, convection-ruled  shells of rotating stars [C. Jordan et al., Mon. Not. R. Astr. Soc. \textbf{252}, Short Communication, 21p–26p (1991)]. Finally, we argue  that the validity of the popular "maximum entropy production principle" [L. M. Martyushev, V. D. Seleznev, Phys. Rep. \textbf{426}, 1 (2006)] is questionable in these systems. 
\end{abstract}



\end{titlepage}
 
\section{The problem}
\label{SEC1}

When considering a physical system which is far from thermodynamic equilibrium and is 'open', i.e. which exchanges either mass, momentum or energy with the external world, it is customary to write the total time derivative $ \dot{S} $ of 
Boltzmann-Gibbs entropy

\begin{equation} \label{entropia}
S = - k_{B} \int f \ln \left( f \right) \mbox{d}V 
\end{equation}

of the system in the form \cite{Glansdorff}:

\begin{equation} \label{bilanciodientropia}
\dot{S} = \Pi - \Phi
\end{equation}

Here and in the following, the word 'entropy' refers to Boltzmann-Gibbs entropy \eqref{entropia} only. We have adopted the notation of \cite{Tomé0}, \cite{Tomé} and \cite{Casas}; 
$ k_{B} $ and 
$ f = f \left( \textbf{x}, t \right) > 0$ 
are Boltzmann's constant and 
the distribution function respectively; 
$ f \left( \textbf{x}, t \right) \mbox{d}V $ 
is the probability of finding the system at the time $ t $ inside the volume element $ \mbox{d}V $ of a $ N$-dimensional state space which is centered at 
$ \textbf{x} = \left( x^{1}, \ldots x^{N} \right)$ and satisfies the normalization condition 
$ \int f \mbox{d}V = 1 $
at all times. For instance, $ x^{1}, \ldots x^{N} $ may be coordinates in $ \mathbb{R}^N $, concentrations of chemical species, etc. 
Moreover, $ \Pi $ and $ \Phi $ are the contributions of physical processes occurring inside the bulk of the system and at the system boundary respectively (some authors denote $ \Pi $ and $ \Phi $ with $ \frac{d_{i}S}{dt} $ and $ -\frac{d_{e}S}{dt} $ respectively; we drop also the dependence on $ \textbf{x} $ and 
$ t $ unless otherwise specified below, for simplicity). 

The quantity $ \Pi $ may include e.g. Joule heating, viscous heating etc. The quantity $ \Phi $ may be related to the exchange of mass, energy etc. with the external world; it is the amount of entropy which escapes away from the system across the system boundaries per unit time and may be $\lesseqgtr 0$, depending on the particular problem. 
It is customary to write $ \Pi $ and $ \Phi $ as a volume integral of a suitably defined entropy production density $ \sigma $ (referred to as 'entropy source strength' in \cite{Williams}) on the system volume and as a surface integral of some suitably defined entropy flow across the system boundary respectively. Here $ \sigma \geq 0 $ at all times everywhere throughout the system because of the second principle of thermodynamics (as for the counterexample provided by \cite{Williams}, see  Sec. \ref{SEC6}). It follows that $ \Pi \geq 0 $ at all times -see e.g. Sec. 49 of \cite{LandauFluidi} for a detailed discussion of this issue in the particular case of a viscous, thermally conducting fluid. 

In contrast with 'closed' (i.e., not open) systems where $ S = \max $ at thermodynamic equilibrium, the relative role of $ \Pi $ and $ \Phi $ in open systems far from thermodynamic equilibrium is far from being assessed even in the neighbourhood of steady states - but for a very restricted class of problems \cite{Martyushev} \cite{MS} \cite{Di Vita2} \cite{Gyarmati}. In particular, a crucial issue is the spontaneous evolution ('relaxation') towards a final ('relaxed') state. 

As for closed systems (where $ \Phi \equiv 0 $ at all times), relaxation is related to a growth of $ S $ towards its maximum value. Thermodynamic equilibrium is the relaxed state; its extremum property $ S = \max $, which implies also $ \Pi = \Phi \equiv 0 $, is related to stability of this state against perturbations. Of course, if we decide to apply a transformation of coordinates: 

\begin{equation} \label{diffeomorfismo}
\textbf{x} \rightarrow \textbf{x}^{'}
\end{equation}

the validity of the extremum property $ S = \max $ remains unaffected, just like the equations of motion involved in Liouville theorem underlying it. This conclusion holds even if the actual value of $ S $ may be not invariant under \eqref{diffeomorfismo} - see Appendix \ref{Proofs2}. Invariance  concerns the extremum property, rather than the actual value of the extremised quantity.

As for open systems (where $ \Phi $ may $ \neq 0 $), boundary conditions may keep the relaxed state (if any exists) far from thermodynamic equilibrium, so that $ \Pi > 0$. The question if a description of stable relaxed states of an open system involves extremum properties while requiring no detailed knowledge of the microscopic dynamics of the system - in analogy with the familiar equilibrium thermodynamics - is still open. Usually, once the extremum property is proven relaxation is described as the evolution of a system which starts from some initial condition inside a suitable attraction basin and ends at the configuration corresponding to the extremum. 

Some researchers have invoked the so-called 'local thermodynamic equilibrium' approximation ('LTE') \cite{Glansdorff}. In a nutshell, LTE means that - although the total system is not at thermodynamic equilibrium - the relationships among thermodynamic quantities within a small mass element are the same as in real equilibrium. LTE leads to the so-called 'general evolution criterion', an exact result concerning the total time derivatives of thermodynamic quantities \cite{GP000}, which rules relaxation in many physical systems - see Ref. \cite{Di Vita2} for a review.

Moreover, it has been proven \cite{Gyarmati} that relaxed states in open systems satisfy extremum properties if some very restrictive assumptions are satisfied, like Onsager symmetry relationships $ L_{pq} = L_{qp} $, where $ L $ is Onsager matrix: these properties are Glansdorff and Prigogine's 'Minimum Entropy Production principle' of suitably constrained minimization of $ \dot{S} $, henceforth referred to as 'MinEP', and its equivalent formulation, Onsager and Machlup's 'least dissipation principle'. The relevant formulation of MinEP depends on the actual boundary conditions. The proof of MinEP relies on the positive sign of scalar quantities like the quadratic form $ \sum_{pq} L_{pq} \dot{X_{p}} \dot{X_{q}}$, with $ X_{p} , X_{q} $ $p$-th and $q$-th thermodynamic force respectively. Scalars are invariant under \eqref{diffeomorfismo}, hence MinEP too is invariant under \eqref{diffeomorfismo}. If such restrictions are dropped, MinEP does not apply \cite{Balescu} \cite{Barbera}. 

Among the many attempts to go beyond LTE and Onsager, many authors have postulated a 'Maximum Entropy Production Principle' ('MEPP') in different forms  \cite{Ziegler} \cite{Dewar} \cite{Beretta1} \cite{VonSpakowski} \cite{Beretta2} \cite{Sawada} \cite{Kirkaldy} \cite{Hill} \cite{Martyushev2} \cite{Sekhar} \cite{Kleidon1} \cite{Kleidon2} \cite{Lorenz} \cite{Virgo} \cite{Yoshida} - for a review, see \cite{Martyushev}. According to MEPP, a stable relaxed state corresponds to a suitably constrained maximum of $ \dot{S} $. To date, and in spite of its present popularity, attempts to derive MEPP rigorously meet no general consensus \cite{Di Vita2}  \cite{Polettini00} \cite{Grinstein} \cite{Polettini0}, as they '\emph{are so far unconvincing since they often require introduction of additional hypotheses, which by themselves are less evident than the proven statement}' \cite{Martyushev} - see also the critical review of \cite{MS}. A way out is to drop \eqref{entropia} altogether, and to postulate - in the framework of the so-called Extended Irreversbiel Thermodynamics - that entropy depends locally not only on temperature, pressure etc. but also on the heat flux and the viscous stress tensor \cite{JouCasas}. However, the results reviewed in \cite{Di Vita2} cast doubt on the very existence of a MEPP-like, general-purpose extremum property of $ \dot{S} $ in far-from-equilibrium, stable, relaxed states; for selected classes of problems, however, validity of variational principles concerning particular contributions to the entropy balance remains possible. For instance, many alleged applications of MEPP seem indeed to be statements concerning $ \Phi $ -see Appendix \ref{MEPPapp} for further discussion.

A different, quite successful approach to non-equilibrium, namely 'stochastic thermodynamics' ('ST'), deals precisely with selected classes of physical systems, namely those systems where a non-equilibrium process occurs which is coupled to 
one (or several) heat bath(s) of constant temperature. ST provides a framework for extending the notions of familiar thermodynamics like work, heat and entropy production 
to the level of individual trajectories across the space state of the system - see the reviews of \cite{Seifert} and  \cite{VanderBroeck}. 
Typically, ST describes systems where a few observable degrees 
of freedom - like the positions of colloidal particles or the gross 
conformations of biomolecules - are in non-equilibrium due to the action of 
possibly time-dependent external forces, fields, flow or unbalanced 
chemical reactants. 
In contrast with the approaches described above, which deal with large  ('macroscopic') systems where fluctuations around mean values may be negligible, ST is uniquely suitable for describing smaller ('mesoscopic') systems where the role of (usually quantum-mechanical) fluctuations is not negligible. In the following, and with a slightly misleading wording, we take the words 'macroscopic', 'classical', 'large' and 'non-degenerate' as interchangeable; in the spirit of our thermodynamic treatment, a rigorous definition of the words 'macroscopic' and 'mesoscopic' is provided in Sec. \ref{SEC6ter} which invokes no information on the detailed mechanics underlying fluctuations.

The time-scale separation between the observable, 
typically slow, degrees of freedom of the system and the unobservable fast 
ones (linked either with thermal baths or with the internal dynamics of the 
system, like e.g. in the case of biopolymers), together with the fact that 
temperatures remain well-defined at all times, allows for a consistent 
thermodynamic description in ST. Internal energy, entropy and free energy 
are well-defined and, in principle, computable for fixed values of the slow 
variables. Usually, the time-scale separation implies that the dynamics becomes
Markovian, i.e., the future state of the system depends only on the present 
one with no memory of the past. If the states are made up by continuous 
variables (like position), the dynamics follows a Langevin equation for an 
individual system and a Fokker-Planck equation for the whole ensemble. This is another difference from the above described approaches, which rely on no detailed description of dynamics. 

Rather than to conditions for stability of relaxed states, ST leads to a large number of exact results - usually referred to as 'fluctuation theorems' ('FTs'). Basically, FTs  are mathematical identities derived
from the invariance of the microscopic dynamics under time-reversal - see e.g. assumption (2) in Sec. 2.1 of \cite{EVANS}. They acquire physical meaning by 
associating their mathematical ingredients with the thermodynamic 
quantities identified within ST. 

The success of invariance-based FTs in the description of some mesoscopic systems is confirmed by experiments. It suggests that even a rigorous non-equilibrium thermodynamics of some selected classes of macroscopic systems should take advantage of some property of invariance - just like in some systems Onsager symmetry follows from invariance of microscopic dynamics under time reversal. Remarkably, a derivation of MEPP from the second principle of thermodynamics actually exists, but only provided that '\emph{the maximum flow is taken as a zero flow }[...] \emph{in practice, this can be realised, e.g. by time/space scaling}' \cite{Martyushev3}. Generally speaking, therefore, a different choice of '\emph{time/space scaling}' seems to lead to violation of MEPP, i.e. MEPP may not be invariant under \eqref{diffeomorfismo}. 
In contrast, some authors \cite{Qian} \cite{Polettini0} \cite{Di Vita JCAT} \cite{Polettini} \cite{invarianceofsteady} suggest that invariance under \eqref{diffeomorfismo} is precisely a requirement for meaningful extremum properties for stable relaxed states of selected, open macroscopic systems. (In the following, by 'invariance' we mean 'invariance under \eqref{diffeomorfismo}'). 

The issue is far from being purely academic. Generally speaking, when it comes to systems which remain confined at all times with a certain region $ \Omega $ of the phase space,  \eqref{diffeomorfismo} does \emph{not} leave the measure of $ \Omega $ unaffected. As a matter of principle, therefore, should an invariant extremum property for stable relaxed states actually exist, it would provide information concerning stability regardless of the size of the system; in other words, a common thermodynamic description of stability would be available for both macroscopic and mesoscopic systems.

A simple, far-reaching corollary follows straightforwardly. To date, the relaxed states are usually assumed to be steady states in the non-equilibrium thermodynamics of macroscopic systems - for an exception concerning MinEP, see e.g. Chap. XV of Ref. \cite{Tykodi}. In contrast, cyclic behaviour plays a central role in ST. In fact, despite the fundamental difference between isothermal engines operating at one temperature and genuine heat engines like thermoelectric devices involving two baths of different temperature, ST provides a common framework based on the representation of entropy production in terms of cycles of the underlying network of states \cite{Seifert}. Accordingly, should an invariant extremum property for stable relaxed states actually exist, it would provide information on the stability of both steady and oscillating relaxed states. A bridge could be established between different problems of stability far from thermodynamic equilibrium.

In this paper we focus our attention on a wide - but well-delimited - set of physical systems and assume that relaxed states exist. Our goal is to find an invariant, necessary condition of stability of these relaxed states. We examine the invariant quantities in the entropy balance \eqref{bilanciodientropia} of the system in Sec. \ref{SEC2} . We discuss the connection between invariance and stability of steady relaxed states and hint at a possible criterion for stability in Sec. \ref{SEC3} ; when discussing invariance, we shall make use of dimensionless quantities only, in order to simplify the maths. Since we are looking for necessary condition of stability, we are free to select the perturbations stability is to be checked against; this allows us to focus our attention on stability against slowly evolving perturbations, and leads therefore to dramatic simplification, as shown in Sec. \ref{SEC4} . In Sec. \ref{SEC5} we take advantage of the results of the previous Sections and show that the criterion hinted at in Sec. \ref{SEC3} is actually the looked-for necessary condition for stability of steady relaxed states, starting from the detailed analysis of a relevant, 'pivotal' problem of plasma physics. No Onsager symmetry is assumed. Generalization to oscillating relaxed states is provided in Sec. \ref{SEC6bis} . We show that invariance of physics allows easy reformulation of the result of Sec. \ref{SEC6bis} into a form which is feasible for both macroscopic and mesoscopic systems in Sec. \ref{SEC6ter} . The following Sections deal with further applications in both macroscopic and mesoscopic systems: Rayleigh-Bénard convection (Sec. \ref{SEC5bis}) in a rotating shell (Sec. \ref{ROSSBY}), a flashing ratchet (Sec. \ref{SEC6}) and a Brownian motor (Sec. \ref{Brown}). Conclusions are drawn in Sec. \ref{SEC7}. Detailed computations are to be found in Appendices \ref{Proofs2}, \ref{CONF1app}, \ref{QUAL}, \ref{CGLEapp} and \ref{RossbyAPP}. Appendix \ref{QUAL} contains also a qualitative discussion of our necessary condition of stability starting from ST. 
 
\section{Looking for invariants in an entropy balance} 
\label{SEC2}

The aim of this Section is to review some useful results \cite{Casas} \cite{Polettini} \cite{Qian2002} concerning the behaviour of \eqref{bilanciodientropia} 
under \eqref{diffeomorfismo} for a quite wide set of physical systems with $ N $ degrees of freedom. These systems are described by the following 'PDE' (partial differential equation) problem:

\begin{equation} 
\label{physics}
\begin{cases}
\dfrac{\partial f}{\partial t} + \nabla \cdot \textbf{J} = 0 
\quad ; \quad
\textbf{J} = \textbf{W} f - D \nabla f \quad ; 
\\ \\
\textbf{n} \cdot \textbf{J} = 0
\quad
\mbox{at the boundary}
\end{cases}
\end{equation}

Here $ \textbf{J} $ is a probability density current with $ N $ components, the quantity $ \textbf{W} $ too has $ N $ components, $ \nabla $ refers to partial derivatives on the $ x^{i}$'s ($ i = 1, \ldots N $), and $ D $ is a positive scalar with $ \nabla D = 0$ and 
$ \frac{\partial D}{\partial t} = 0$, which plays the role of a diffusion coefficient in a space with $ N $ dimensions, so that \eqref{physics} acts as a simple balance equation for $ f $. We invoke no information concerning the detailed physical nature of $ \textbf{W} $ and $ D $.  
The boundary condition corresponds to a system which remains confined at all times inside a region $ \Omega $ - with measure $ V = L^N $ - of the state space,  
$ \textbf{n} $ being the unit normal vector of the boundary surface of $ \Omega $. Here and in the following, we compute all integrals in $ \mbox{d}V $ on $ \Omega $ and denote with 
$ <\emph{y}> \equiv V^{-1} \int \emph{y} f \mbox{d}V$ the average of the generic quantity 
$ \emph{y} $ on $ \Omega $. 

Any physical system undergoing a Markovian evolution dictated by a Fokker-Planck equation in thermal contact with an environment (say, a thermal bath) provides us \cite{Polettini} with an example of a system described by \eqref{physics}; generalization is discussed in \cite{Casas}. Another example of physical system described by \eqref{physics} is a single macromolecule \cite{Qian2002} in a fluid, which is composed by $ N_{at} $ atoms and whose configuration is represented at any time by a point $ \textbf{x} $ in a space with $ N = 3 \cdot N_{at} $ dimensions (the fundamental equation describing the evolution of the distribution function of the system, namely equation (5) of \cite{Qian2002}, is a slight generalisation of \eqref{physics} - the main difference is that $ D $ is a tensor). Finally, if $ N = 3 $, the system is made of many particles with the same mass and we identify $ \textbf{W} $ with a velocity (so that $ f $ and $ \textbf{J} $ are $ \propto $ the mass density and the mass current density respectively), then in the limit of negligible $ D $ the PDE \eqref{physics} describes a system where mass is conserved with no net source and is exchanged nowhere across the boundary with the external world.

For the purpose of our discussion, we provide some preliminary information about \eqref{diffeomorfismo}. Quite generally, we assume the $ x^{i}$ 's to be curvilinear coordinates in a $ N-$dimensional Riemannian variety (like e.g. $ \mathbb{R}^N $) with state-space volume element $ \mbox{d}V $. We denote with $ a^{i} $ and $ a_{i} $ the i-th controvariant and covariant component of the generic $ N-$dimensional quantity $ \textbf{a} $ respectively ($ i, j, k = 1, \ldots N $ here and in the following). Moreover, we denote the Jacobian 
$ \det \left( \frac{\partial x^{i'}}{\partial x^{k}} \right) $ of \eqref{diffeomorfismo} 
with $ \Lambda$ . For simplicity, we assume \eqref{diffeomorfismo} to be an orientation-preserving diffeomorphism, so that $ \Lambda > 0$. Note that \eqref{diffeomorfismo} leaves time unaffected. Finally, we denote with $ y^{'} $ the value of the generic quantity $ y $ after the transformation. 

First of all, it is shown in Appendix \ref{Proofs2} that:

\begin{equation} \label{Pi}
\Pi = 
\int \frac{k_{B}}{D} \frac{\vert \textbf{J} \vert^2}{f} \mbox{d}V
\end{equation}

\begin{equation} \label{Phi}
\Phi = 
\int \frac{k_{B} \left( \textbf{J} \cdot \textbf{W} \right)}{D}  \mbox{d}V
\end{equation}

As for \eqref{Pi}, basically a particular case of equation (19) of \cite{Tomé0}, it retrieves both equation (12) of \cite{Casas} and equation (7) of \cite{Qian2002} provided that $ D \propto T$ in agreement e.g. with Einstein's relation ($ T $ is a scalar quantity, hence it is invariant). Indeed, both $ \nabla T $ and $ \frac{\partial T}{\partial t} $ are supposed to vanish in both \cite{Polettini} and \cite{Qian2002}, just like $ \nabla D $ in our treatment. This assumption is justified because of the interaction with a thermal bath in \cite{Polettini}, and is explicitly stated in \cite{Qian2002}, where $ D \propto T $. Finally, in Sec. \ref{SEC4} we are going to discuss the behaviour of $ \Pi $ in the limit of negligible $ D $. 

As for \eqref{Phi}, it is a particular case of equation (20) of \cite{Tomé0}. If $ \textbf{W} $ is 'conservative', i.e. if $ \textbf{W} = \nabla \vartheta$ where the scalar quantity $ \vartheta $ is a differentiable function of $ \textbf{x} $, then our boundary condition makes $ \Phi $ to vanish after integration by parts in steady state (where $ \dot{S} = 0$ and $ \nabla \cdot \textbf{J} = 0 $); \eqref{bilanciodientropia} implies therefore $ \Pi = 0 $. (An alternative definition of a conservative $ \textbf{W} $ is $ \nabla \wedge \textbf{W} = 0 $, where the $i$-th component of $ \nabla \wedge \textbf{a} $ is $ \varepsilon_{ijk} \frac{\partial a_{j}}{\partial x_{k}}, i,j,k = 1, \ldots N $ for the generic $ \textbf{a} $). Accordingly, if $ \textbf{W} $ is conservative then thermodynamic equilibrium is the only possible steady state, and far-from-equilibrium steady states are possible only for other boundary conditions - think e.g. of a net amount of current flowing across the boundary. Far-from-equilibrium steady states with the boundary condition of \eqref{physics} are possible only if $ \textbf{W} $ is not conservative. Not surprisingly, the latter statement is false when $ D \propto T $ and $ \nabla T \neq 0 $ \cite{Tomé} \cite{Polettini}, since $ \nabla D \neq 0 $ in this case; the question is briefly assessed in Sec. \ref{SEC4}.

Now, it comes to the discussion of invariance. Physical intuition dictates that \eqref{bilanciodientropia} and \eqref{physics} take the same form before and after the transformation \eqref{diffeomorfismo}. As for the quantities involved in \eqref{physics}, it is shown in Appendix \ref{Proofs2} that this requirement of invariance implies:

\begin{equation} \label{trasformazionediD}
D^{'} = D
\end{equation}

\begin{equation} \label{densitascalare}
f^{'} = \dfrac{f}{\Lambda}
\end{equation}

\begin{equation} \label{densitavettoriale}
J^{' i} = \dfrac{1}{\Lambda} \dfrac{\partial x^{' i}}{\partial x^{k}} J^{k} 
\end{equation}

\begin{equation} \label{trasformazionediA}
W^{'}_{i} = \dfrac{\partial x^{' k}}{\partial x^{i}} 
\left[
W_{k} - D \dfrac{\partial \ln \left( \Lambda \right)}{\partial x^{k}}
\right]
\end{equation}

As for the quantities involved in \eqref{bilanciodientropia}, it is shown in Appendix \ref{Proofs2} that the values of both $ S $, $ \dot{S} $ and $ \Phi $ are not invariant; in contrast, $ \Pi $ is invariant:

\begin{equation} \label{invariante2}
\Pi^{'} =
\Pi
\end{equation}

Accordingly, the only invariant quantities in \eqref{bilanciodientropia} are $ \Pi $ and $ D $. The roles of $ \Pi $ and $ D $ are discussed in Sec. \ref{SEC3} and Sec. \ref{SEC4} respectively. We discuss in Appendix \ref{CONF1app} a further aspect of  \eqref{invariante2}. 

In spite of its simplicity, \eqref{densitascalare} has a far-reaching consequence. For a  sufficiently well-behaved $ f $, the theorem of the mean allows us to rewrite the invariant normalization condition  
$ \int f \mbox{d}V = 1 $ as 
$ 1 = f \left( \textbf{x}_{M} \right) \cdot V $
where $ \textbf{x}_{M} $ belongs to $ \Omega $. Together, invariance of the latter relationship and equation \eqref{densitascalare} lead to: $ \frac{V'}{V} = \frac{f \left( \textbf{x}_{M} \right)}{f ' \left( \textbf{x}_{M} ' \right)}  = \Lambda $, a quantity which may  differ from 1. This means that $ V $ is usually not invariant. It follows that if an invariant criterion of stability exists, then it holds regardless of the actual value of $ V $; in particular, it may hold for both macroscopic and mesoscopic systems, as anticipated in \ref{SEC1} . Sec. \ref{SEC6ter} presents further, in depth-discussion of the issue.

For future reference, here we discuss also the particular case of rescaling. By 'rescaling' we mean a diffeomorphism \eqref{diffeomorfismo} where
$ \frac{\partial x^{' i}}{\partial x^{k}} = \Lambda \delta_{ik} $, $ \delta_{ik} = 0 $ if $ i \neq k $ and $ \delta_{ik} = 1 $ otherwise, and $ \nabla \Lambda = 0 $. Equation \eqref{physics} is invariant under rescaling; indeed, if $ \nabla \Lambda = 0 $ then \eqref{densitascalare}, \eqref{densitavettoriale} and \eqref{trasformazionediA} ensure that $ f $ transforms just like $ \nabla $, that $ \textbf{J} $ is invariant, and that $ \textbf{W} $ transforms like a vector.

\section{Invariance and stability}
\label{SEC3}

The aim of this Section is to assess the impact of \eqref{invariante2} on the stability of steady relaxed states of a system described by \eqref{physics} against small perturbations. Under mild assumptions, the system evolves - with a typical time-scale $ \tau \propto D^{-1} $ - towards a relaxed state where the current $ \textbf{J} $ has no sinks and sources, i.e. $ \nabla \cdot \textbf{J} = 0 $. The relaxed state coincides with thermodynamic equilibrium when $ \Pi = 0 $, hence $ \textbf{J} = 0$, while non-equilibrium relaxed states are characterized by non-vanishing currents that circulate in the system's state space. The words 'small perturbation' refer to a perturbation with small measure in the Riemannian variety of the $ x^{i}$'s. The word 'stability' refers to the fact that if we apply a small perturbation to a steady state then the perturbation relaxes back to the initial state; $ \frac{\partial}{\partial t} $ (which is zero before the perturbation) takes a non-zero value with a well-defined sign (possibly oscillating in time), and after a while relaxes back to zero. 

To start with, we observe that the set $ \textit{M} $ of transformations \eqref{diffeomorfismo}, \eqref{trasformazionediD}, \eqref{densitascalare}, \eqref{densitavettoriale} and \eqref{trasformazionediA} maps the space $ \lbrace \textbf{x}, D, f, \textbf{J}, \textbf{W} \rbrace $ onto itself. By construction, $ \textit{M} $ leaves \eqref{physics} unaffected. Then, $ \textit{M} $ maps solutions of \eqref{physics} onto solutions of \eqref{physics}. Moreover, $ \textit{M} $ is both reflexive, symmetric and transitive (as it can be shown with the help of both chain rule and the well-known properties of Jacobians after cumbersome but straigthforward algebra), i.e. $ \textit{M} $ is a relationship of equivalence. Consequently, $ \textit{M} $ establishes a partition of the set of the solutions of \eqref{physics} in equivalence classes; each class is unambiguously labelled by a value of $ \Pi $ because of \eqref{invariante2}. Furthermore, $ \textit{M} $ leaves $ t $ unaffected. Then, $ \textit{M} $ maps steady states onto steady states. Different steady states of the same equivalent class correspond to different choices of $ \textbf{x}, D, f, \textbf{J} $ and $ \textbf{W} $, which are mapped onto each other by $ \textit{M} $. Outside steady states, invariance of \eqref{physics} ensures that the sign of $ \frac{\partial}{\partial t} $ is also invariant. 
Accordingly, $ \textit{M} $ maps stable steady states onto stable steady states. 
Finally, different choices of $ \textbf{x}, D, f, \textbf{J} $ and $ \textbf{W} $ may even correspond to different physical systems, as far as that the latter satisfy the same equation \eqref{physics}. For example, if $ \textbf{W} $ is a velocity then it is ruled by the equations of motion, which in turn contain the forces acting on the systems. Different systems are subject to different forces, and the behaviour of $ \textbf{W} $ changes accordingly. However, the proof of \eqref{invariante2} relies on no invariance of the equations of motion. Correspondingly, $ \textit{M} $ may map different physical systems onto each other even if \eqref{physics} remains valid. It follows that $ \textit{M} $ can map the stable steady state (if any exists) of a physical system which satisfies \eqref{physics} onto the stable steady state of another physical system which satisfies \eqref{physics}, both steady states having the same value of $ \Pi $. The basins of attraction of stable states may be different in different physical systems, as the equations of motion may be different; but as far as these equations include \eqref{physics}, our argument remains unaffected. For example, $ M $ maps thermodynamic equilibria ($ \Pi = 0 $) onto thermodynamic equilibria ($ \Pi ' = \Pi = 0 $).

Now, in Sec. \ref{SEC5} we describe one particular ('pivotal') physical system which satisfies \eqref{physics} and which enjoys the following property: a necessary condition for the stability of a steady state against small, slowly evolving perturbations is that this state satisfies:

\begin{equation} \label{variational}
\Pi  = \min
\end{equation}

under suitable constraints. We stress the point that \eqref{variational} differs from both MinEP and MEPP, which involve $ \dot{S} $ and not just $ \Pi $. Small perturbations of a stable steady state raise $ \Pi $ above its value in steady state; after a while, $ \Pi $ relaxes back to the initial value. Minimization in \eqref{variational} is constrained by the requirement that the solutions of \eqref{variational} are also solutions of the equations of motion in steady state. While steady states correspond to solutions of the equations of motion in steady state, stable steady state solutions satisfy also \eqref{variational}. 

Indeed, the following facts occur: a) when we apply \eqref{diffeomorfismo} then invariance of \eqref{physics} forces $ \textbf{x}, D, f, \textbf{J} $ and $ \textbf{W} $ to get transformed under $ \textit{M} $; b) all steady states obtained from the stable, relaxed state of the pivotal system via $ \textit{M} $ have the same, constant value of $ \Pi $, which is a minimum in the pivotal system; c) all unsteady, perturbed states obtained from the unsteady, perturbed states of the pivotal system have the same, time-dependent value of $ \Pi $, which is larger than the minimum value of $ \Pi $ in the pivotal system; d) $ \textit{M} $ leaves the sign of $ \frac{\partial}{\partial t} $ unaffected, so that relaxing states are mapped onto relaxing states. 

Together, these facts lead to the conclusion that once the pivotal system is given, the same necessary criterion for stability \eqref{variational} holds for all physical systems which satisfy \eqref{physics} and which are obtained from the pivotal system via  \eqref{diffeomorfismo}. Indeed, when one of such systems relaxes back to a stable steady state after a perturbation, then the value of $ \Pi $ in the relaxing state and in the relaxed state respectively are equal to the corresponding values of the pivotal system, where \eqref{variational} holds. 

We conclude that invariance of \eqref{bilanciodientropia} and \eqref{physics} implies \eqref{variational} for steady, stable, relaxed states of systems described by \eqref{physics} provided that we find a pivotal system. Noteworthy, we invoke no Onsager symmetry and 
no detailed knowledge of the physical nature of $ \textbf{x}, f, \textbf{J} $ and $ \textbf{W} $. A qualitative argument which invokes no pivotal system is provided in Appendix \ref{QUAL} .

\section{Slow diffusion}
\label{SEC4}

The aim of this Section is to assess the impact of \eqref{trasformazionediD} on the stability of relaxed states. Since we are looking for a necessary condition of stability, we are free to select the perturbations which stability is to be checked against. We focus our attention on slowly evolving perturbations \cite{Di Vita2} and denote with $ \tau $ the typical time-scale of their relaxation in the following. For large $ \tau \propto D^{-1} $ we assume that ('slow diffusion approximation'):

\begin{equation} \label{slowdiffusion}
\vert D \nabla f \vert \ll \vert \textbf{W} f \vert
\end{equation}

According to \eqref{trasformazionediD}, \eqref{trasformazionediA} and \eqref{invariante2},  \eqref{slowdiffusion} is invariant. 

The impact of \eqref{slowdiffusion} on stability is no new topic in the literature. Sec. VII of \cite{Qian2002} links \eqref{variational} and \eqref{slowdiffusion} in the particular case of the relaxation of a physical system made of two coupled subsystems at different temperatures, where Joule heating is the only dissipative process in the bulk of each subsystem, so that $ \Pi $ is just equal to $ \frac{1}{T} $ times the Joule heating power $ \int P_{J} \mbox{d}V $ ($ P_{J} $ Joule power density). In each subsystem it is assumed that \eqref{physics}, $ \nabla T = 0 $ and $ D \propto T $ hold. It turns out that if \eqref{slowdiffusion} holds then the relaxed state minimizes the amount of heat produced per unit time in the bulk by irreversible phenomena. This implies that \eqref{variational} holds, as $ \nabla T = 0 $ in each subsystem. In contrast, if $ \vert D \nabla f \vert \gg \vert \textbf{W} f \vert $ then we are near thermodynamic equilibrium, i.e. in Onsager’s regime, and MinEP holds.

More generally, \eqref{slowdiffusion} leads to the consequences listed below. 

\begin{itemize}
\item Even in the extreme case $ D \rightarrow 0 $ equation \eqref{Pi} gives 
$ 0 < \Pi < \infty $ provided that 
$ \vert \textbf{J} \vert \propto O \left( \sqrt{D} \right) $. Physically, a finite, positive amount of entropy is produced in the bulk per unit time even if $ D $ provides just a higher-order correction to the equations of motion like \eqref{physics}. Even if seemingly rather artificial, the latter statement applies to many physical systems where dissipative phenomena like the diffusion represented by $ D $ in \eqref{physics} allow relaxation, but are not necessarily invoked when it comes to a detailed description of the relaxed states itself. For example, friction damps small oscillations of a pendulum around its position of stable equilibrium until the pendulum stops, but provides no information about the final position itself. This may be true even if no conservative forces occur, like e.g. in the case of stable equilibria of magnetically confined plasmas; indeed, these equilibria are described by Grad-Shafranov equation of dissipation-free magnetohydrodynamics ('MHD') even if the relaxation processes leading to them is ruled by dissipation \cite{Freidberg}. This discussion plays a crucial role in Appendix \ref{CGLEapp} below.
\item \emph{Per se}, \eqref{slowdiffusion} does not imply $ D \equiv 0 $; rather, it is an useful simplification for the investigation of slowly evolving perturbations. This issue turns out to be relevant in Sec. \ref{SEC6} below.
\item Admittedly, this simplification comes at a price. When computing $ \Pi $ in a relaxed state, we cannot straightforwardly rely on equation \eqref{Pi}; we need rather an explicit expression for $ \Pi $ as provided e.g. by the equation of motions of the particular physical system under investigation. This is e.g. the case of the problems discussed in both Secs. \ref{SEC5} and \ref{SEC5bis} below. 
\item Furthermore, \eqref{slowdiffusion} implies that $ D $ provides just a higher-order correction to the dynamics of the system implies that the same holds for $ \nabla D $ too. It follows that our results apply even if $ \nabla D \neq 0 $ within a negligible error. 
The case $ D \propto T $ is of particular interest. In this case, if $ D \propto T $ then both $ T $ and $ \nabla T $ affect \eqref{physics} through $ D $, and have therefore no impact in the slow diffusion approximation. According to \eqref{slowdiffusion}, we expect therefore our results below to apply to both systems with $ \nabla T = 0 $ and $ \nabla T \neq 0 $. In particular, such systems may belong to the same equivalence class even if the equations of motion may differ; for example, gravitating fluids with $ \nabla T \neq 0 $ may exhibit convection, which can be absent when $ \nabla T = 0 $. We anticipate here that $ \nabla T = 0 $ in the pivotal system in Sec. \ref{SEC5}; all the same, according to our discussion we shall be able to apply \eqref{variational} even to $ \nabla T \neq 0 $ systems in Secs. \ref{SEC5bis} and \ref{SEC6} .
\item The quantity $ D $ describes diffusion; in order to maintain a stable steady state satisfying \eqref{slowdiffusion}, it is therefore required that the amount of heat produced per unit time within the bulk of the system is also small; hence $ \Pi $ too is small. For example, the macroscopic fluids dealt with in Secs. \ref{SEC5} and \ref{SEC5bis} below dissipate only weakly. In the general case, \eqref{physics}, \eqref{Pi} and \eqref{slowdiffusion} give 
$ D = \frac{k_{B}}{\Pi} L^N  < \vert \textbf{W} \vert^2 >$ and 
$ \vert \nabla_{y} \ln \left( f \right) \vert \approx 
\frac{\vert \textbf{W} \vert}{< \vert \textbf{W} \vert^2 >} \frac{\Pi}{k_{B} L^{N-1}} $ 
%
%
where $ \textbf{y} \equiv \frac{\textbf{x}}{L} $ and
$ \nabla_{y} \equiv \frac{\partial}{\partial \textbf{y}} = L \nabla$. If $ \Pi $ is small enough, it follows that:

\begin{equation} \label{slowlyvaryingf}
\vert \nabla_{y} \ln \left( f \right) \vert \ll 1
\end{equation}

All the way around, \eqref{slowlyvaryingf} implies \eqref{slowdiffusion}.
\item Relationships \eqref{physics}, \eqref{trasformazionediA} and \eqref{slowdiffusion} make $ \textbf{W} $ to transform like a vector and allow us to write:

\begin{equation} 
\label{physicssimplified}
\dfrac{\partial f}{\partial t} + \nabla \cdot \left( \textbf{W} f \right) = 0
\quad ; \quad
\textbf{n} \cdot \textbf{W} = 0
\quad
\mbox{at the boundary} 
\end{equation}

where $ \textbf{J} - \textbf{W} f $ is either zero or the gradient of an harmonic function with vanishing normal derivative at the boundary, so that \eqref{slowlyvaryingf} implies:

\begin{equation} 
\label{C}
\nabla \wedge \textbf{J} = f \nabla \wedge \textbf{W} 
\end{equation}

All the way around, if $ \textbf{J} $ satisfies \eqref{C} then \eqref{slowlyvaryingf} allows $ \textbf{J} $ to satisfy \eqref{physics} 
up to an additive gradient of an harmonic function.

\end{itemize}

\section{A pivotal system}
\label{SEC5}

The aim of this Section is to describe a pivotal system. Here and in the following, we limit ourselves to the $ N = 3 $ case and we take $ \mathbb{R}^3 $ as the state space. We are going to show that the electrons in the pinch of a Dense Plasma Focus ('DPF')  \cite{Bernard} \cite{Soto} form a pivotal system. 
 
\begin{figure}[!h]
\centering
\includegraphics[scale=0.6]{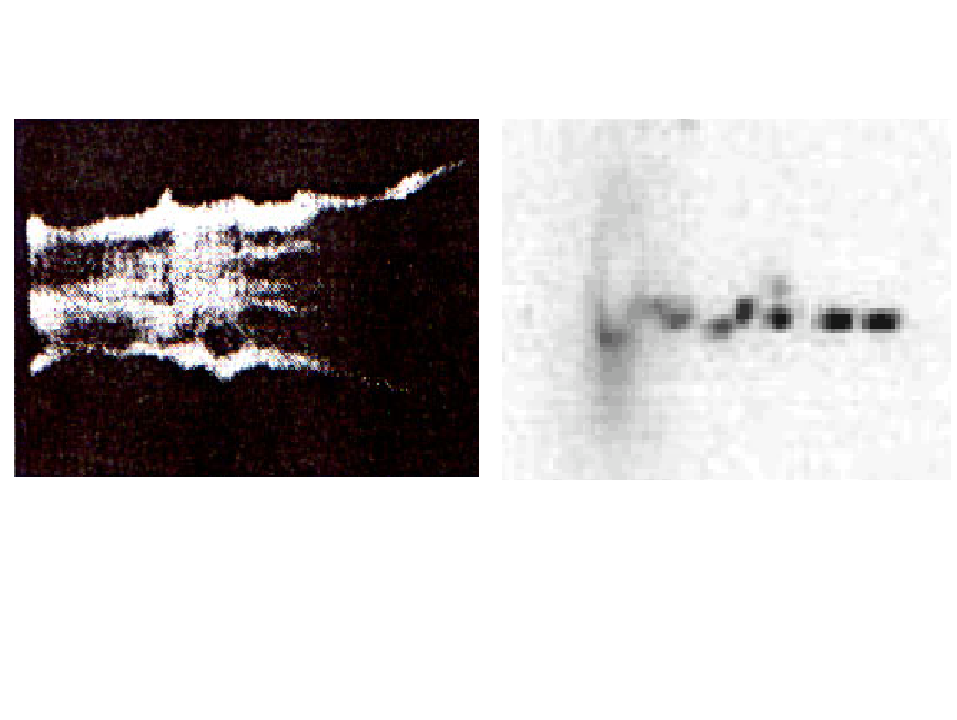}
\caption{\textit{X-ray pinhole images of the filamentary structure (left, positive image) \cite{Herold}  and of plasmoid-like structure \cite{Jakubowski} in various DPF pinches (right, negative image). The structure can be resolved as a sequence of small, intense sources of  radiation inside a more diffused source. Vertical size of the picture $ 5 $ mm, filament thickness $ \approx $ plasmoid diameter in the range $ 0.1 - 0.3 $ mm \cite{BostickPrior}. According to \cite{Di Vita}, the volume of the dissipating region in the pinch shrinks to small filaments and plasmoids, in order to minimize $ \int P_{h} d\mbox{V} $; $ P_{h} \approx P_{J} $ in plasmoids, unlike filaments.}}
\label{PFfilamentplasmoid}
\end{figure}

A Dense Plasma Focus ('DPF') is a kind of discharge in which a high-pulsed voltage is applied to a gas between electrodes, generating a short-duration, high-density plasma region. The pinch observed at the final stage of a well-behaved DPF discharge is made of a dense, hot plasma. This plasma is an example of classical (i.e., non-degenerate), electrically neutral, viscous, resistive, electrically conducting, magnetised fluid, where the electrons carry electric current and the (much heavier) ions ensure charge neutralization, so that the ratio of ion and electron densities is fixed everywhere at all times (we refer to both densities as to 'particle density' below). The inner structure of DPF pinch spontaneously exhibits typical, complex patterns, both in the form of filaments \cite{Herold} and of plasmoids \cite{Jakubowski} - see Fig.~\ref{PFfilamentplasmoid} .

The life-time of the DPF pinch is $\approx 10^{-6}$ s $ \gg $ the decay time-scale $\approx 10^{-9}$ s predicted by resistive MHD, so we are allowed to speak of a 'steady state'. Accordingly, $ \Pi $ is the sum of four terms, due to Joule heating, viscous heating, heat transport (due to conduction and radiation) and particle diffusion \cite{Landau0}. The contributions of Joule heating and viscous heating are the volume integrals of $ \frac{1}{T} $ times the Joule heating power density $ P_{J} $ and the viscous heating power $ P_{V} $ density respectively. Both radiation and turbulence ensure efficient transport of heat across the pinch, so we may safely assume both $ \nabla T = 0$ and vanishing gradient of particle density; the contributions of heat transport and particle diffusion to $ \Pi $ are therefore negligible. Correspondingly, \eqref{variational} is equivalent to simultaneous minimization of Joule heating power $ \int P_{J} d\mbox{V} $ (Kirchhoff's principle \cite{Jaynes}) and viscous heating power $ \int P_{V} d\mbox{V} $ (Kortweg and Helmholtz' principle \cite{Lamb}), i.e. to minimization of $ \int P_{h} d\mbox{V} $ where  $ P_{h} \equiv P_{J} + P_{V}$. Now, simultaneous validity of Kirchhoff's and Kortweg and Helmholtz' principles is precisely a necessary condition of stability of steady, relaxed states of viscous, electrically conducting fluids against small, slowly evolving perturbations \cite{Di Vita2} \cite{Di Vita} \cite{Di Vita3}. As discussed below, the equations of motion in steady state constrain the minimisation. Then, we may say that \eqref{variational} is a necessary condition of stability for the relaxed state of this system. Indeed, it has been shown in Sec. 8 of \cite{Di Vita} that the above minimization leads to: 

\begin{equation}
\label{London}
\nabla^{'} \wedge \textbf{j}^{'} = - \mu_{0}^{-1} g^{2} \textbf{B}^{'}
\end{equation}

in the DPF pinch. Here 
$ \textbf{j}^{'} $ is the electric current density,
$ \mu_{0} = 4 \cdot \pi \cdot 10^{-7} T \cdot A^{-1} \cdot m $,
$ g = \mu \left( \mu_{0} \cdot f \right)^{\frac{1}{2}} $,
$ \mu $ is the electric charge/mass ratio of the electrons
(in the notation of \cite{Di Vita} the quantity $ w = - \mu $ is utilised), 
and  $ \textbf{B}^{'} = \nabla{'} \wedge \textbf{A}^{'} $ is the magnetic field 
with vector potential $ \textbf{A}^{'} $. Both $ \textbf{j}^{'} $ and $ \textbf{B}^{'} $ are functions of the position vector 
$ \textbf{x}^{'} $ inside our fluid, and 
$ \nabla^{'} $ refers to partial derivatives on the components of $ \textbf{x}^{'} $. Physically, 
\eqref{London} plays the role of a London equation where London depth is equal to the collisionless skin depth of electrons. It describes successfully \cite{Di Vita} the filamentary structure of electric currents flowing across the DPF pinch \cite{Herold} provided that the Hartmann number $ Ha $ 
(the ratio of electromagnetic force to the viscous force) 
is lower than a certain threshold $ Ha_{c} $ (in a turbulent plasma of pure hydrogen and with current $ I $, $ Ha_{c} = 4.4 \cdot 10^3 \sqrt{\frac{I \left( MA \right)}{T \left( KeV \right)}}$ \cite{DiVitaHartmann}). (Experiments show that the impact of velocity, which $ P_{V} $ depends on, is definitely not negligible in DPF \cite{Pavez} filaments). Ultimately, it is this success which justifies \eqref{variational} in this system. The case $ Ha > Ha_{c} $ is discussed below and corresponds to plasmoids.

We want to show that we are actually dealing with a pivotal system. To this purpose, we  show that the electrons in the DPF pinch satisfy \eqref{physics} and \eqref{slowdiffusion}, or, equivalently, \eqref{slowlyvaryingf} and \eqref{physicssimplified}. Then, we link \eqref{C} - itself a consequence of \eqref{slowlyvaryingf} - and \eqref{London}. 

For the moment, we assume \eqref{slowdiffusion} to hold; this assumption is justified below. We apply a rescaling 
$ \textbf{x}^{''} = M_{tot}^{-1} \textbf{x} $ 
to \eqref{physics}, where $ M_{tot} $ is the total mass of DPF electrons in the pinch and 
$ \Lambda = M_{tot}^{-1} $. As anticipated in Sec. \ref{SEC1} , we are going to utilize dimensionless quantities. In dimensionless form, for example, $ M_{tot} $ is the total number of electrons. Relationships \eqref{trasformazionediD} and \eqref{slowdiffusion} ensure that the impact of $ D $ remains negligible. Then, \eqref{densitascalare} and \eqref{densitavettoriale} ensure that $ f^{''} $ and $ \textbf{J}^{''} $ are $ M_{tot}^{-1} $ times the electron mass density and mass density flow respectively. The PDE in \eqref{physicssimplified} is just the mass balance in a system - like the electrons in a  DPF pinch - where no net source of mass exists. The boundary condition in \eqref{physicssimplified} is discussed below. 

As for \eqref{slowdiffusion}, the facts that $ \nabla T = 0$ and that $ f \propto $ mass density, together with 
equations (5.2) and (6.1) of \cite{Di Vita}, ensure that the pressure gradient $ \propto \nabla \left( T f \right) \approx T \nabla f $ is $\approx 0$; in turn, this justifies  \eqref{slowlyvaryingf}, hence \eqref{slowdiffusion}. 
Finally, $ \nabla f \approx 0 $ agrees \cite{LandauFluidi} with the fact that the contribution of mass transport to $ \Pi $ is negligible, as anticipated. 

Moreover, \eqref{London} is just a particular case of \eqref{C} provided that 
$ \textbf{x}^{'} = \textbf{x}^{''} $ 
(the corresponding $ \textit{M} $ is an identity) and that we take
$ \textbf{j}^{'} = \mu \textbf{J}^{''} $
and
$ \textbf{A}^{'} = - \mu^{-1} \textbf{W}^{''}$. 
Together, the latter relationship and the fact that $ \textbf{W} $ transforms like a vector ensure that the vector potential transforms like a vector, as expected.
Gauge invariance allows addition of the gradient of an harmonic function $ \chi $ to the vector potential to leave physics unaffected in steady state. As expected from Sec. \ref{SEC4}, dissipation provides no information on the relaxed state; indeed, \eqref{London} contains no information concerning dissipation, and the latter plays a minor role in the momentum balance. 
 
We are left to discuss both the role of the boundary condition in \eqref{physicssimplified} and the constraints on minimization in \eqref{variational}. For simplicity of notation, we drop all superscripts below, so that $ \textbf{x} $ is the position vector inside the fluid, etc. 

As for the boundary condition $ \textbf{n} \cdot \textbf{W} = 0 $, it has been assumed in \cite{Di Vita} that the interaction of the system with fields due to external sources is much weaker than the interaction among currents internal to the pinch. In our language, this is equivalent to neglect the impact of the external world represented by non-vanishing net flux of charge carriers across the boundaries of the system, i.e. to $ \textbf{n} \cdot \textbf{J} = 0 $. 

Admittedly, the virial theorem of MHD forbids the existence of magnetically confined plasmas without interaction with external currents. However, Sec. 6 of \cite{Di Vita} shows that the impact of such interaction is comparable to the impact of dissipation, which is considered to be weak in agreement with \eqref{slowdiffusion}. Even so, and in spite of $\nabla T = 0$, if $ \textbf{B} \neq 0 $ then $ \textbf{A} \propto \textbf{W} $ is not conservative and a steady state far from thermodynamic equilibrium is possible; $ \textbf{B} $ is the '\emph{curvature}' in the language of \cite{Polettini}. In contrast with the 1D systems of \cite{Pol22}, however, our systems are 3D. 

As for the constraints on the minimization \cite{Di Vita}, they are provided by the mass balance quoted above, Maxwell's equation of electromagnetism and Navier-Stokes equation with a bulk force density corresponding to the electromagnetic force. For an electrically neutral, magnetized fluid Maxwell's equation reduce to Gauss' and Ampère's law, while the electromagnetic term in the momentum balance corresponds to the density of Lorenz force. The viscous term in Navier-Stokes equation is as small as the impact of interaction with external currents, as stated above. Moreover, the actual value $ T_{\mbox{boundary}} $ of $ T $ throughout the system provides minimization 
$ T \left( \textbf{x} \right) = T_{\mbox{boundary}} $ 
with the further constraint throughout the system \cite{Di Vita}, as $ \nabla T = 0 $.

We stress the point that - even if electrons are the electric charge carriers in a DPF pinch - the analysis of \cite{Di Vita} relies on the particular nature of the electric charge carriers nowhere. In other words, validity of \eqref{variational} does not depend on the actual value of $ \mu $. Accordingly, we allow $ \mu $ to take arbitrary real values below. If $ \mu \neq 0 $, \eqref{London} makes the electron velocity to be the sum of a potential term and a quantity $ \propto \mu $ times a rotational field; in the limit $ \mu \rightarrow 0$, a potential flow is retrieved, the structure on the small spatial scale of the electron collisionless skin depth is lost and \eqref{London} reduces to the identity $0 = 0$. Unlike \eqref{C}, however, \eqref{London} follows from Kirchhoff' and Kortweg and Helmholz' principles, which in turn agree with \eqref{variational} for $ \nabla T = 0 $ only; accordingly, relaxed states with $ \nabla T \neq 0 $ may have non-potential flows even if $ \textbf{B} = 0 $ - see e.g. Secs. \ref{SEC5bis} and \ref{Brown} below.

Admittedly, there is still a weak point in our discussion. In Ref. \cite{Di Vita}, the proof of Kirchhoff's and Korweg and Helmholz' principles in the DPF pinch relies on LTE. Thus, one could think that validity of \eqref{variational} is limited to systems where LTE holds. In order to show that this is not the case, let us discuss what happens if $ Ha > Ha_{c} $. According to the same constrained minimization invoked above, it turns out that the relaxed configuration is a plasmoid \cite{Jakubowski}, with spheroidal geometry and $ P_{h} \approx P_{J} $ as $ P_{J} \propto \vert \textbf{j} \vert^2 $, $ \textbf{j} \propto \textbf{B} $ and $ \vert \textbf{B} \vert $ is very large ($ \lesssim 10^5 $ T) \cite{Bernard}; moreover, a plasmoid is approximately described by Taylor's principle of minimum magnetic energy with fixed magnetic helicity \cite{Taylor00} , and no velocity affects Taylor's principle - see \cite{Di Vita} , \cite{DiVitaHartmann} and Refs. therein. Correspondingly, $ \textbf{A}^{'} \neq - \mu^{-1} \textbf{W}^{''}$ and \eqref{London} follows from \eqref{C} no more, even if the velocity of charge carriers ($ \propto$ the electric current density) may still have non-zero vorticity. 
Since $ P_{h} \approx P_{J} \propto $ the electrical resistivity $ \eta_{\Omega} $,  we expect the $ P_{h}$-minimizing relaxed state to correspond - all other things being equal - to some minimum of $ \eta_{\Omega} $ or of some monotonically increasing function of $ \eta_{\Omega} $. Indeed, this is precisely what happens with plasmoids, according to the independent analysis leading to equations (1)-(4) of \cite{Comisso}. We may take this fact as a confirmation of our discussion, provided that we take into account the following three facts. Firstly, it starts from investigation of plasmoids in astrophysics, not in DPF.  Secondly, it does not invoke LTE; validity of \eqref{variational} is therefore not limited to systems where LTE holds. Finally, it relies on a postulate in the form of a Fermat-like principle concerning the whole relaxation of a reconnection-affected magnetized fluid towards a plasmoid; as such, the final outcome depends on the initial condition (more precisely, on the initial inverse aspect ratio of the current sheet of interest): as a result, a whole class of relaxed plasmoids are obtain, each of them corresponding e.g. to a given value of embedded magnetic flux, an approximately constant quantity for weakly dissipating plasmas. 

In conclusion, electrons in a DPF pinch are an example of a classical fluid with viscous and Joule heating, negligible mass diffusion (i.e., \eqref{slowlyvaryingf} holds), no net mass source and negligible mass flow across the boundary (i.e., \eqref{physicssimplified} holds), and $ \nabla T = 0 $. If a steady, stable state of this fluid exists, then it satisfies also the variational problem \eqref{variational}. Then, we identify this fluid with the pivotal system. We drop the restriction $ \nabla T = 0 $ altogether in Sec. \ref{SEC5bis}. In particular, if $ \mu \neq 0 $ (the pivotal system is magnetized), $ \nabla T = 0 $ and $ Ha > Ha_{c} $ then equation \eqref{C} follows from \eqref{slowlyvaryingf} and reduces to the London-like equation \eqref{London} describing a filamentary pattern of electric currents. If $ \mu \neq 0 $, $ \nabla T = 0 $ and $ Ha < Ha_{c} $ then a plasmoid-like structure arises. 

\section{Oscillating vs. steady}
\label{SEC6bis}

Until now, we have assumed that fluctuations occur on time-scales much shorter than the time-scale $ \tau $ of the perturbation, in order to allow us to get rid of such fluctuations through suitable time averaging. Nothing is said about the detailed frequency spectrum of the fluctuations. In this Section we discuss what happens if the fluctuations oscillate in time with period $ \tau_{P} \ll \tau $. We denote -as usual by now- the generic physical quantity by $ \emph{y} $, write: 

\[\emph{y} \left( t \right) = 
\emph{y}_{0} +
\emph{y}_{10} \cos \left( \frac{2 \pi t}{\tau_{P}} \right) +
\emph{y}_{20} \cos \left( \frac{2 \pi t}{\tau} \right) + 
\ldots \]

and introduce the time-averaging
$ \widehat{\left[\emph{y}\right]} \left( t \right) \equiv 
\frac{1}{\tau_{aux}} \int_{t}^{t + \tau_{aux}} d\mbox{t} \emph{y} \left( t' \right) $ 
with
$ \tau >> \tau_{aux} >> \tau_{P} $,  
so that 
$ \widehat{\left[\emph{y}\right]} = \emph{y}_{0} +
\emph{y}_{20} \cos \left( \frac{2 \pi t}{\tau} \right) + 
\ldots 
+ O \left( \frac{\tau_{aux}}{\tau} \right) $
and 
$ \widehat{ \left[ \frac{d \emph{y}}{d t} \right] } = 
\frac{d \widehat{\left[\emph{y}\right]}}{d t} 
+ O \left( \frac{\tau_{P}}{\tau_{aux}} \right) $. The same result holds if we replace  
$ \cos \left( \frac{2 \pi t}{\tau} \right) $
with a generic function of time evolving on a typical time-scale $ \tau $. 
Accordingly, time-averaging just erases all contributions of the dynamics on the fast time-scale $ \tau_{P} $ and allows us to replace $ \emph{y} $ with $ \widehat{\left[\emph{y}\right]} $ everywhere while leaving all relationships involving $ \emph{y} $ and its time derivative unaffected -up to a small error 
$ \approx O \left( \frac{\tau_{aux}}{\tau} , \frac{\tau_{aux}}{\tau} \right) $ at least. 
To the best of author's knowledge, this argument has been invoked for the first time in Chap. XV of \cite{Tykodi} for systems satisfying MinEP. 
Up to the same error, we are therefore allowed to say that a steady, stable (unstable) state is a configuration where the value of $ \emph{y} $ undergoes fluctuations with time-scale $ \tau_{P} $ and where $ \widehat{\left[\emph{y}\right]} $ is stable (unstable) against small perturbations evolving on a time-scale $\tau >> \tau_{P}$. 

We retrieve physical examples of such configurations when investigating e.g. systems described by \eqref{physics} with $\textbf{W}$ periodic function of time. These systems may relax to oscillating states, where both amplitude and period $ \tau_{P} $ of the oscillations depend on the detailed structure of $\textbf{W} \left( t \right)$. We may ask if these oscillating states are stable against perturbations evolving on time-scales much longer than $ \tau_{P} $. Now, both \eqref{bilanciodientropia}, \eqref{Pi}, \eqref{Phi} and \eqref{internal0} remain valid even if $\textbf{W}$ depends on $ t $ - as it can be shown with step-by-step repetition of the proof of equation (10) in \cite{Casas}. Maps $ M $ still leave $ t $ unaffected, and \eqref{invariante2} still holds. It follows that 
$ \left( \widehat{\left[\emph{y}\right]} \right) ' = \widehat{\left[\emph{y}'\right]} $; 
if $ \emph{y} = \Pi $ then 

\begin{equation}
\label{invariant00}
\widehat{\left[\Pi\right]}' = \widehat{\left[\Pi\right]}
\end{equation}

i.e. $ \widehat{\left[\Pi\right]} $ too is invariant. Of course, the instantaneous value of $ \Pi $ in a stable relaxed state coincides at all times with $ \widehat{\left[\Pi\right]} $ if and only if the stable relaxed state is a steady state, i.e. in the limit $ \tau_{P} \rightarrow \infty $. In the general case, the whole argument concerning the pivotal system still holds after replacement of $ \Pi $ with $ \widehat{\left[\Pi\right]} $. Thus, \eqref{variational} and \eqref{invariant00} lead to the following necessary condition for stability:

\begin{equation}
\label{variationaloscillating0}
\widehat{\left[\Pi\right]} = \min
\end{equation}

If $\textbf{W}$ becomes constant, $ \tau_{P} \rightarrow \infty$ and we retrieve \eqref{variational} as expected; we limit therefore ourselves to the discussion of \eqref{variationaloscillating0} below. After time-averaging of both its sides, \eqref{internal0} gives $ \widehat{\left[\Pi\right]} = \widehat{\left[\Phi\right]} $ because 
$ \widehat{\left[\dot{S}\right]} = 0 $ as $ S $ too is periodic in time; $ \widehat{\left[\Phi\right]} $ may differ from zero even if $ \widehat{\left[\textbf{W}\right]} = 0 $ because of correlations.

\section{Mesoscopic vs. macroscopic}
\label{SEC6ter}

Remarkably, the invariance property \eqref{invariante2} depends on no actual value of $ V $, and the same class of equivalence may contain both macroscopic systems (like e.g. the pivotal one) and mesoscopic systems, i.e. $ M $ may map macroscopic and mesoscopic systems onto each other. We may think e.g. of a rescaling with a suitably defined value of $ \Lambda $, so that $ V $ changes as discussed in Sec. \ref{SEC2} . If we apply $ M $ to a macroscopic system and, as a result, $ V $ gets shrinked too much, the linear dimension of the resulting $ V $ may become of the same order of magnitude of the De Broglie length of the resulting system, and the latter is definitely mesoscopic. All the way around, we could start from a mesoscopic system satisfying \eqref{physics} (and also \eqref{slowdiffusion}, as far as stability against slow perturnations is concerned) and obtain a macroscopic fluid with no net mass source, no mass flow across the boundary and  negligible particle diffusion (in agreement with \eqref{slowdiffusion}).  As far as $ \Pi $ is invariant, the problems of stability of a star and of a molecular motor may therefore be dual to each other, and share a common solution. Of course, a star produces an immensely larger amout of entropy per unit time; but the actual value of $ \Pi $ is not relevant to the problem of stability, just its minimum property counts. (Analogously, stability of thermodynamic equilibrium requires just maximisation of $ S $, and does not depend on the actual value of $ S $). 

\begin{figure}[!h]
\centering
\includegraphics[scale=0.4]{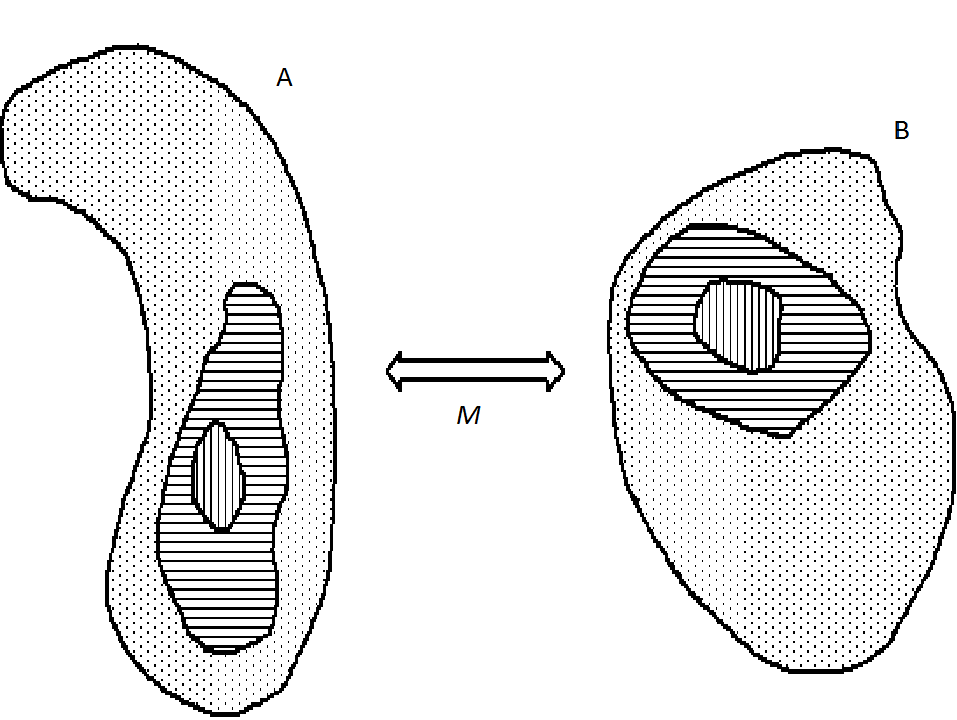}
\caption{\textit{Let two distinct physical systems A and B satisfy \eqref{physics} . A and B are displayed as separated groups of nested sets on the left and the right side of the figure respectively. A set $ M $ (displayed by the central arrow in the figure) of transformations given by \eqref{diffeomorfismo}, \eqref{trasformazionediD}, \eqref{densitascalare}, \eqref{densitavettoriale} and \eqref{trasformazionediA} maps A and B onto each other while leaving \eqref{physics} unaffected. The only physical quantities entering the entropy balance which are preserved by $ M $ are $ \Pi $ and $ D $. Both A and B may be either macroscopic or mesoscopic. In order to investigate thermodynamic properties of relaxed states, we assume that for each system the set of solutions of \eqref{physics} (dotted region) includes a subset of relaxed configurations (horizontal lines). The actual state the system relaxes to depends on the detailed dynamics of the system, the initial conditions etc. In turn, the subset of relaxed configuration may further include a subset (vertical lines) of configurations which are both relaxed and stable against slow perturbations, evolving on some time-scale $ \tau $. For each system, relaxed states may either be steady states of oscillating states with period $ << \tau $. Since we limit ourselves to discuss necessary conditions of stability against slow perturbations, we may invoke the slow diffusion approximation \eqref{slowdiffusion}, which means that the impact of $ D $ on \eqref{physics} is small - again, a condition invariant under $ M $. Any classical fluid with no net mass source, no mass flow across the boundary and negligible diffusion of mass is an example of macroscopic system satisfying \eqref{physics} and \eqref{slowdiffusion}. As far as both A and B satisfy \eqref{physics}, $ M $ maps relaxed states onto relaxed states, and stable ones onto stable ones, regardless of the detailed dynamics of A and B; for example, A and B can be a magnetized macroscopic fluid and a mesoscopic system respectively. Furthermore, if a 'pivotal' system (say, A) exists where a necessary condition for stability of a relaxed state is constrained minimization of $ \overline{\widehat{\left[\Pi\right]}} $, the constraints being given by the equations of motion, then the same condition holds for B; even if the equations of motion of A and B are different, minima of $ \overline{\widehat{\left[\Pi\right]}} $ in A correspond to minima of $ \overline{\widehat{\left[\Pi\right]}} $ in B. We show that a pivotal system exists, and apply the resuling conditions for stability to different systems, both macroscopic and mesoscopic. Finally, a qualitative discussion in Appendix \ref{QUAL} suggests that the actual existence of a pivotal system is not even required.}}
\label{provaschema}
\end{figure}

We may therefore ask ourselves if, given an oscillating relaxed state in a mesoscopic system, there is some reformulation of \eqref{variationaloscillating0} which applies to the stability of such state. We need this reformulation for two reasons: a) a mesoscopic system which belongs to an equivalence class may still satisfy \eqref{slowlyvaryingf} and \eqref{physicssimplified}, but $ f $ and $ \textbf{J} $ are no more to be given the meaning of mass density and mass flow respectively; b) the impact of fluctuations should be explicitly taken into account. If it exists, such reformulation may hold for both macroscopic and mesoscopic systems. This is the topic of this Section.

To start with, we stress the point that the proof of \eqref{variationaloscillating0} relies on \eqref{Pi}. In turn, the proof of \eqref{Pi} relies on the positiveness of $ \Pi $. In average, this positiveness follows from the fact that at the trajectory level, realizations with positive entropy production are exponentially more likely than the corresponding realizations with negative entropy production \cite{VanderBroeck}. The probability of observing a 'reverse' behaviour, i.e. an entropy production opposite to that dictated by the second law of thermodynamics, decreases exponentially as the system size increases. Thus, careful discussion of the averaging procedure is required when discussing mesoscopic systems. 

The quantity $ \widehat{\left[\Pi\right]} $ depends on the path $ \textit{C} $, i.e. the actual trajectory followed by the system in phase space starting from the configuration at $ t = 0 $ and arriving at the configuration at the time $ t $ (with no loss of generality, we suppose time averaging is computed on the time interval from $ 0 $ to $ t $ here and in the following.). We define \cite{Tomé} \cite{Crooks} the path ensemble average 
$ \overline{\textit{y}} \equiv \sum_{\textit{C}} \textit{y} \left( \textit{C} \right) \textit{p} \left( \textit{C} \right)$ 
of the path-dependent, generic quantity $ \textit{y} \left( \textit{C} \right)$ where the sum is over the set of all paths connecting all possible initial and final states and the probability $ \textit{p} \left( \textit{C} \right) $ of the path $ \textit{C} $ depends on both the detailed dynamics of the system and the probability of the initial state. The smaller the system, the more likely the probability of observing a 'reverse' behaviour, the more relevant the distinction between $ \widehat{\left[\Pi\right]} $ and $ \overline{\widehat{\left[\Pi\right]}} $. In contrast, we may safely write 
$ \overline{\widehat{\left[\Pi\right]}} = \widehat{\left[\Pi\right]}$ in macroscopic systems, where 'reverse' behaviour is extremely unlikely. Accordingly, and regardless of the detailed mechanics of fluctuations, we may provide the following definitions, just as anticipated in Sec. \ref{SEC1} : 

\begin{itemize}
 \item the relationship 
 $ \overline{\widehat{\left[\Pi\right]}} - \widehat{\left[\Pi\right]} = 0 $ 
defines a macroscopic system;
 \item the relationship 
$ \overline{\widehat{\left[\Pi\right]}} - \widehat{\left[\Pi\right]} \neq 0 $ defines a mesoscopic system.
 \end{itemize} 

Now, by definition a system which is initially in a stable state relaxes back to its initial configuration after a small perturbation has been applied. This means that -after a sufficiently long time at least- both the $ \textit{C} $'s and the corresponding $ \textit{p} \left( \textit{C} \right) $'s are unaffected. This property holds as far as the system is stable. Since we are looking for necessary conditions of stability, we are free to choose the quantity to be perturbed in order to check stability. With no loss of generality, let us consider a quantity $ \textit{y} $ which is a function of $ \textbf{x} $ such that the transformation of coordinates $ \textbf{x} = \left( x^{1}, \ldots x^{N} \right) \rightarrow \textbf{x}' = \left( x^{1}, \ldots x^{N-1}, \textit{y} \right)$ is a diffeomorfism with positive Jacobian $ \Lambda $. Accordingly, any small perturbation 
$ d \textit{y} $ of $ \textit{y} $ on a time-scale $ \tau $ is equivalent to a small perturbation $ d \Lambda $ of $ \Lambda $ on the same time-scale. The perturbation $ \Lambda \rightarrow \Lambda + d \Lambda$ corresponds to an infinitesimal $ M $ which maps stable states onto stable states. If we apply it to a stable state, then the $ \textit{p} \left( \textit{C} \right) $'s remain unaffected and  
$ \left( \overline{\textit{y}} \right)' = \overline{\textit{y}'} $. This relationship remains valid even if we repeat the transformation $ \Lambda \rightarrow \Lambda + d \Lambda$ again and again, so that we may take it to be valid even for a finite $ \Lambda$. 
If $ \textit{y} = \widehat{\left[\Pi\right]} $, then we obtain 
$ \left( \overline{\widehat{\left[\Pi\right]}} \right)' = \overline{\widehat{\left[\Pi\right]}'} = \overline{\widehat{\left[\Pi\right]}}$ where the last equality follows from \eqref{invariant00}. If the system we start from is  macroscopic, then 
$ \overline{\widehat{\left[\Pi\right]}} = \widehat{\left[\Pi\right]}$. 
Stability implies \eqref{variationaloscillating0}, hence:

\begin{equation}
\label{variationaloscillating}
\overline{\widehat{\left[\Pi\right]}} = \min
\end{equation}

As anticipated, the crucial point is that -even if the starting system is macrosopic- the final system needs not to be macroscopic; \eqref{variationaloscillating} is a generalization of \eqref{variationaloscillating0}, valid for both mesoscopic and macroscopic systems. As for mesoscopic systems, an open question in ST research is whether the probability distributions of work, heat and entropy production can be grouped into "universality classes" characterized, e.g., by the asymptotics of such distributions, and, if yes, which specific features of a system determine this class \cite{Seifert}. By now, it is temptative to identify each "universality class" with the equivalence class labeled by the value of $ \overline{\widehat{\left[\Pi\right]}} $ which the mesoscopic system belongs to. As for macroscopic systems, $ \overline{\widehat{\left[\Pi\right]}} = \widehat{\left[\Pi\right]}$ and we retrieve \eqref{variationaloscillating0} as expected; we limit therefore ourselves to the discussion of \eqref{variationaloscillating} below. 

It is clear, by now, that a suitable choice of the value of $ M_{tot} $ allows a rescaling to map any mesoscopic system which obeys \eqref{physics} and \eqref{slowdiffusion}, i.e. \eqref{slowlyvaryingf} and \eqref{physicssimplified}, onto a pivotal system for some value of $ \mu $, so that the system belongs to an equivalence class and its stable relaxed states satisfy \eqref{variationaloscillating}. As for \eqref{physics}, $ f $ and $ \textbf{J} $ are to be given the meaning of mass density and mass flow respectively no more, in contrast with what occurs in macroscopic systems, as anticipated in Sec. \ref{SEC6ter}. As for \eqref{slowdiffusion}, it is equivalent to take into account slowly evolving perturbations only, as anticipated in Sec. \ref{SEC4}.

The proof of \eqref{variationaloscillating} relies on the existence of a pivotal system, through \eqref{variationaloscillating0} and \eqref{variational}. If, furthermore, $ \nabla T = 0 $ everywhere and $ \frac{\partial T}{\partial t} = 0 $ at all times (as it happens e.g. when the system is in contact with just one heat reservoir at temperature $ T $), then an independent, qualitative confirmation of \eqref{variationaloscillating} exists - see Appendix \ref{QUAL} - which invokes no pivotal system and shows its agreement with ST ; see Fig.~\ref{provaschema} for an overall sketch of our results.

\section{Rayleigh-Bénard convection}
\label{SEC5bis}

Until now we have dealt with the pivotal system, where $ \nabla T = 0 $. Generally speaking, if a macroscopic, $ \nabla T = 0 $ system is a fluid with Joule and viscous heating in relaxed, steady state then \eqref{variationaloscillating} agrees with Kirchhoff's and Kortweg and Helmholtz' principles invoked in Sec. \ref{SEC5}. 
Remarkably, however, the arguments leading to \eqref{variationaloscillating} invoke $ \nabla T = 0 $ nowhere. In particular, we invoke the argument presented in Sec. \ref{SEC4} : if $ D \propto T $ then $ T $ affects \eqref{physics} through $ D $ only; then, in the slow diffusion approximation $ T $ leaves the validity of \eqref{variationaloscillating} unaffected - and the same holds for 
$ \nabla T $. In the following, we are dealing with applications of \eqref{variationaloscillating} to systems where $ \nabla T \neq 0$. In these systems, and in in contrast with the pivotal system, the transport of energy is not so strong to flatten $ \nabla T $ . For example, different boundaries of the same system may be in thermal contact with heat reservoirs at different temperatures \cite{Tomé} \cite{Polettini} \cite{Qian2002}. In such systems, far-from-equilibrium relaxed states are possible even if $ \textbf{W} $ is conservative.  

The aim of this Section is to discuss a class of macroscopic systems, namely fluids with Joule and viscous heating where Rayleigh-Bénard convection occurs, where $ \nabla T \neq 0$ and with negligible mass diffusion, no net mass sources and no mass flow across the boundary. Convection is triggered when the Rayleigh number $ Ra $ - basically, a dimensionless version of $ \vert \nabla T \vert $ -  exceeds a threshold $ Ra_{c} $. We discuss mesoscopic systems with $ \nabla T \neq 0$ in Sec. \ref{SEC6}.

Physically, Joule heating and viscous heating are related to the transport of different quantities, namely electric charge and momentum respectively. Correspondingly, they provide $ \sigma $ with different, additive contributions and different thermodynamic flows. Heat transport provides a further, additive contribution with a further thermodynamic flow. Entropy production in our fluid is therefore a 'compound process' \cite{MS}, and validity of MEPP is questionable - see Appendix \ref{MEPPapp} .

We show in Appendix \ref{CGLEapp} that a $ Ra \lesssim Ra_{cr} $ steady, relaxed, possibly unmagnetized fluid belongs to the same class of equivalence of a particular pivotal system. More to the point, we can map a Rayleigh-Bénard problem with its own boundary condition (Dirichlet, Neumann or a linear combination of the two) in a steady, relaxed fluid on the verge of convection onto the steady, relaxed filamentary configuration of a DPF, and vice-versa. Differences between the dynamics of the two systems do not matter; what is relevant here is that such mapping is possible. The proof takes into account the fact - hinted at in Sec. \ref{SEC4} - that the description of a stable relaxed state involves no dissipation; the latter affects rather the description of the relaxation process, which is very slow in the slow diffusion approximation and may therefore safely being described as a succession of steady states. 

This result, together with the above-quoted fact that the impact of non-zero $ \nabla T $ on stability is negligible in the slow diffusion approximation, allows us to invoke \eqref{variationaloscillating0}, i.e. the version of \eqref{variationaloscillating} for macroscopic systems, as a necessary condition of stability for relaxed states in Rayleigh-Bénard convection problems. The problems of the stability of the relaxed steady states of the DPF filaments and the $ Ra \lesssim Ra_{cr} $ fluid are dual to each other. In the latter fluid $ \textbf{W} $ plays a role of a fluid velocity, and $ \nabla \wedge \textbf{W} \neq 0$ as convection cells rotate, say, with period $ \tau_{P} $. A far-from-equilibrium relaxed state is possible as $ \nabla T \neq 0$. 

Rayleigh-Bénard convection is the result of the simultaneous occurrence of gravitation (which, \textit{per se}, leaves the entropy balance \eqref{bilanciodientropia} unaffected) and of a temperature jump across a fluid layer. Both are absent in the pivotal fluid of Sec. \ref{SEC5}. In contrast, these systems share the following, common features: a) mass balance holds with no net mass source and no mass flow across the boundaries, i.e. \eqref{physicssimplified} holds with $ f $ mass density, $ \textbf{J} $ mass density flow; b) viscous and Joule dissipation are the only heating processes of the bulk (Joule heating plays a role in magnetized fluid), while the impact of particle diffusion on entropy production is negligible - in agreement with \eqref{slowdiffusion} and \eqref{slowlyvaryingf} - so that $ \Pi $ includes a new term due to heat conduction. The entropy balance of a small mass reads:

\begin{equation} \label{dsdt}
\rho T \dfrac{ds}{dt} = P_{h} 
- \nabla \cdot \textbf{q}
\end{equation}

where $ \textbf{q} $ is the heat flow due to conduction and radiation \cite{Di Vita2}. 
If no mass flows across the boundary, Reynolds' transport theorem \cite{Smirnov} and  integration of both sides of \eqref{dsdt} on the volume of the system gives back \eqref{bilanciodientropia} with 
$ \Pi = \int \frac{P_{h}}{T} d\mbox{V} + 
\int \textbf{q} \cdot \nabla \left( \frac{1}{T} \right) d\mbox{V} $ and 
$ \Phi = \int \frac{\textbf{q}}{T} \cdot \textbf{n} da $ 
after invoking Gauss' theorem of divergence and after integration by parts. Both 
$ \int \frac{P_{h}}{T} d\mbox{V} $ and $ \int \textbf{q} \cdot \nabla \left( \frac{1}{T} \right) d\mbox{V} $ are $ \geq 0 $ and the latter is an increasing function of the former; then, \eqref{variationaloscillating0} implies minimization of 
$\widehat{\left[\int\textbf{q}\cdot\nabla\left(\frac{1}{T}\right)d\mbox{V}\right]}$. We invoke this minization in order to discuss a Rayleigh-Bénard problem below. We are going to retrieve some results, which include the well-known variational principle of Chandrasekhar \cite{Chandrasekhar} at $ Ra \approx Ra_{cr}$. Our argument is a slightly modified version of the argument presented in Sec. VI.B of \cite{Di Vita2}. Here the novelty is that we invoke LTE no more; we have replaced it with the assumption of no mass flow across the boundary.

It is clear, by now, that our pivotal system acts just as a Trojan horse which allows us to smuggle the stability criterion \eqref{variationaloscillating} into the world of the relaxed solutions of \eqref{slowlyvaryingf} and \eqref{physicssimplified}. Even in the framework of the slow diffusion approximation, the relevance of \eqref{variationaloscillating} does not depend on the detailed physics. Rather, it follows from the very existence of a partition of the set of the solutions of \eqref{physics} in equivalence classes - in other words, on the invariance of $ \Pi $ under \eqref{diffeomorfismo}. 

A far-reaching corollary follows. Both gravitation and non-zero $ \nabla T $ leave and \eqref{dSdt} and \eqref{dsdt} unaffected. Then, we may apply our discussion above to the relaxed state of any classical fluid with Joule and viscous heating, negligible mass diffusion, no net mass source and negligible mass flow across the boundary. Regardless of $ \nabla T $, stability of the relaxed states of this fluid requires \eqref{variationaloscillating}. The detailed dynamics of the system appears in the constraints of minimization, and depend on the particular problem.

Now, let us come back to Rayleigh-Bénard convection. 
The contribution of heat conduction to $ \Pi $ increases with increasing $ \vert \nabla T \vert $, for not too large $ \vert \nabla T \vert $ at least (the precise meaning of 'not too large' is given below); it vanishes only if $ \nabla T = 0 $ everywhere - see Sec. 49 of \cite{LandauFluidi} and Sec. 66 of \cite{Landau0}. Accordingly,  \eqref{variationaloscillating0} implies minimization of 
$ \widehat{\left( \vert \nabla T \vert \right)} $.

As for the constraints on minimization, we want to retrieve the constraints given by the equations of motion and by $ T \left( \textbf{x} \right) = T_{\mbox{boundary}} $ in the limit of vanishing $ \nabla T $. It can be shown \cite{Di Vita2} that the correct choice of constraints for the $ \nabla T \neq 0$ which satisfy this condition is given by the equations of motion and by a given value of $ \int P_{h} d\mbox{V} $. 
The equations of motion include the mass balance with no net mass source and the Navier-Stokes equation; in an electrically conducting fluid, Maxwell's equation of electromagnetism are also included, and the density of Lorenz force appears in the Navier Stokes equation. No mass flows anywhere across the boundaries. After time-averaging, the energy balance obtained from the equations of motion ensures that the time-averaged total heating power is equal to the time-averaged mechanical power supplied by buoyancy (with density $ P_{b} $); this relationship acts as a constraint on minimization. 

Under such constraint, minimization of $ \widehat{\left[ \vert \nabla T \vert \right]} $ retrieves the  result derived in Secs. 33 and 43 of \cite{Chandrasekhar} from the analysis of the equations of motion near the onset of Rayleigh-Bénard's convection cells in both unmagnetised and magnetised fluids. Indeed, the author keeps the value of $ \nabla T $ fixed, so that the quantity actually minimized is $ \vert \nabla T \vert $:

\begin{equation}
\label{Chandrasekhar}
\vert \nabla T \vert = \min
\qquad
\mbox{with the constraint}
\widehat{ \left[ \int P_{h} d\mbox{V} \right] } = \widehat{ \left[ \int P_{b} d\mbox{V} \right] }
\end{equation}

Since $ \vert \nabla T \vert \propto Ra $ in the problems solved in \cite{Chandrasekhar}, the words 'not too large' above mean $ Ra \approx Ra_{cr} $. 

\section{Convection in a rotating shell}
\label{ROSSBY}

Some consequences of \eqref{Chandrasekhar} are discussed in \cite{Di Vita2}. Here we apply \eqref{Chandrasekhar} to the description of the relaxed state of a macroscopic, rotating spherical shell 
- with inner radius 
$ R_{\min} $, outer radius $ R_{\max} = R_{\min} + \Delta R$,  
thickness $ \Delta R > 0 $, 
volume $ V_{s} = \frac{4 \pi}{3} \left( R_{\max}^3 - R_{\min}^3 \right) $ and rotation period $ \emph{P} $ - 
of a viscous, electrically conducting fluid where a non-vanishing temperature jump drives convective cells with typical linear size 
$ L_{c} $ rotating with period 
$ \tau_{P} $ and typical rotation velocity 
$ u_{c} = \frac{L_{c}}{\tau_{P}}$. We are allowed to invoke \eqref{Chandrasekhar} as  rotation affects \eqref{bilanciodientropia} unaffected. This is e.g. the case of the subphotospheric, convection-ruled zone of rotating stars \cite{Priest}, where typical values of the Rossby number 
$ Ro \equiv \frac{\emph{P}}{\tau_{P}} $ \cite{Jordan000} are $ 0.1 \leq Ro < \infty $ \cite{Bercik}. 
This convection zone is usually supposed to be a spherically symmetric shell where convection and dynamo effect \cite{Moffatt} sustain a temperature gradient and a magnetic field \textbf{B} respectively \cite{Belvedere}. In the following, we derive a scaling law for \textbf{B} from \eqref{Chandrasekhar}. 

\begin{figure}[!h]
\centering
\includegraphics[scale=0.6]{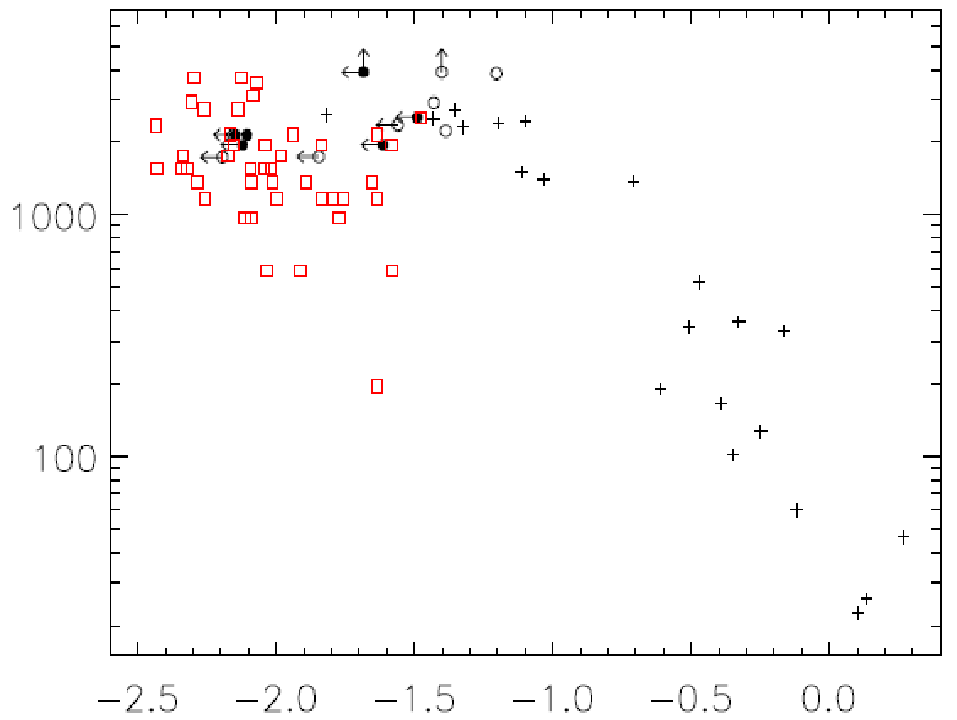}
\caption{\textit{Magnetic field (Gauss) vs. $ \log_{10} Ro $, from Fig. 19 of \cite{Reiners} . A tight dependence between magnetic field and $ Ro $ is clearly observed in sun-like (crosses) and early- to mid-M type stars (other dots). The result is
that magnetic activity is rising with decreasing $ Ro $ as long as $ Ro \geq 0.1 $. At $ Ro \approx 0.1 $, activity saturates and does not grow further with decreasing $ Ro $. This behavior is interpreted as increasing dynamo efficiency with faster rotation in the $ Ro \geq 0.1 $ regime where rotation is not yet dominating convection. At fast rotation ($ Ro \leq 0.1 $) the dynamo reaches a level of saturation that cannot be exceeded even if the star is spinning much faster.}}
\label{BvslogRo}
\end{figure}

The reciprocity principle of variational calculus - see Sec. IX.3 of \cite{Elsgolts} - makes the solution of \eqref{Chandrasekhar} to solve also the variational problem:

\begin{equation}
\label{Chandrasekhar00}
\widehat{ \left[ \int P_{h} d\mbox{V} \right] } = \max
\qquad
\mbox{with the constraint of fixed}
\vert \nabla T \vert
\end{equation}

Further discussion requires two auxiliary assumptions, to be dropped below. Firstly, we neglect viscosity, i.e. we replace $ P_{h} $ with 
$ P_{J} = \frac{{\vert \textbf{E} \vert}^2}{\eta_{\Omega}}$ where $ \textbf{E} $ is the electric field and 
$ \eta_{\Omega} = \eta_{\Omega} \left( T \right) $ is supposed to be a scalar. Faraday's law 
$ \nabla \wedge \textbf{E} +  \frac{\partial \textbf{B}}{\partial t} = 0 $
gives 
$ \vert \textbf{E} \vert \approx \vert \textbf{v} \vert \vert \textbf{B} \vert $ 
where
$ \vert \textbf{v} \vert \approx u_{c} $. 
Secondly, we focus our attention on the case 
$ \Delta R \ll \frac{T}{\vert \nabla T \vert} $, so that we make a small error if we replace 
$ \eta_{\Omega} \left( T \right) $ 
with 
$ \eta_{*} \equiv \eta_{\Omega} \left( \frac{R_{\max} + R_{\min}}{2} \right) $ 
everywhere inside the shell. For example, according to \cite{Belvedere} solar dynamo occurs in a narrow layer 
$ 0.65 R_{\bigoplus} < R < 0.70 R_{\bigoplus} $ between the convection and the radiation zone ($ R_{\bigoplus} $ is the solar radius). 

As a further, more fundamental hypothesis, we assume small electromagnetic fluctuations, so that we may neglect contributions of dynamo to $ P_{J} $ \cite{Boozer} and write 
$ < \vert \textbf{B} ^2 \vert > \approx < \vert \textbf{B} \vert > ^2 $ 
in the relaxed configuration of the shell. Thus, we write 
$ P_{h} = P_{J} = \frac{{\vert \textbf{E} \vert}^2}{\eta_{\Omega}} \approx 
\frac{u_{c}^2 \vert \textbf{B} \vert ^2}{\eta_{*}} $ 
with 
$ u_{c} = \frac{L_{c}}{\tau_{P}} = 
\frac{L_{c} Ro}{\emph{P}}$, hence:

\begin{equation}
\label{Chandrasekhar11}
\widehat{ \left[ \int P_{h} d\mbox{V} \right] } \approx 
\dfrac{V_{s} L_{c}^2 Ro^2 < \vert \textbf{B} \vert > ^2}{\eta_{*} \emph{P}^2}
\end{equation}

Cowling's theorem \cite{Moffatt} forbids occurrence of dynamo in two-dimensional systems: convection cells in the dynamo-relevant zone are therefore likely to exhibit non-trivial three-dimensional structure. Azimuthal periodicity provides us with an upper bound $ L_{c} \leq L_{c \max} = u_{c} \emph{P} $. Then, maximization in \eqref{Chandrasekhar00} requires that we replace $ L_{c} $ with $ L_{c \max} $  in \eqref{Chandrasekhar11}, and the constraint in \eqref{Chandrasekhar} gives:

\begin{equation}
\label{Chandrasekhar1111}
< \vert \textbf{B} \vert > \approx \dfrac{\emph{F}_{R}}{Ro}
\end{equation}

where 
$ \emph{F}_{R} \equiv \frac{1}{u_{c}} \sqrt{\frac{\eta_{*} 
\widehat{ \left[ \left( \int P_{b} d\mbox{V} \right) \right] }}{V_{s}}} $. Here we have implicitly assumed that maximization affects just $ L_{c} $ and not directly $ \vert \textbf{B} \vert $, which in turn means that we neglect possible saturation of dynamo at very large rotation velocity, i.e., very low $ Ro $. If, furthermore, the centrifugal force is not too strong, then the dependence of $ \emph{F} $ on $ Ro $ is weak. In fact, both 
$ u_{c} $, 
$ \eta_{*} $, 
$ V_{s} $ and $ P_{b} $ depend rather - through 
$ R_{\min} $, $ T \left( R_{\min} \right) $, 
$ R_{\max}$ and $ T \left( R_{\max} \right) $ - 
on the rotation-independent chemical composition and mass of the star, as well as on its global balances of energy and momentum, i.e. the hydrostatic balance of gravity and pressure gradient, etc. The same holds for the value of $ \vert \nabla T \vert $ in \eqref{Chandrasekhar00}. Moreover, $ \emph{F}_{R} $ is likely to be the same for the stars of the same spectral class, so that \eqref{Chandrasekhar1111} leads to:

\begin{equation}
\label{Rossby}
< \vert \textbf{B} \vert > \propto Ro^{- \emph{x}}
\qquad
\emph{x} = 1
\qquad
\mbox{when comparing the stars of the same spectral class}
\end{equation}

In Appendix \ref{RossbyAPP} we discuss some independent estimates of 
$ \emph{x} $ which confirm \eqref{Rossby} and which do not rely on the auxiliary assumptions referred to above. Indeed, observations \cite{Jordan000} for F and G/K dwarf stars are in agreement with 
$ \vert \textbf{B}_{c} \vert \propto Ro^{- \emph{x}} $, where
$ \textbf{B}_{c} $ is the coronal field and 
$ 0.78 < \emph{x} < 1.9 $. A review \cite{Reiners} which includes both Sun-like and M-type stars shows that \eqref{Rossby} fits data as far as $ - 1 < \log_{10} Ro $, while the above-neglected dynamo saturation occurs at $ \log_{10} Ro < - 1 $  - see Fig.~\ref{BvslogRo} .
 
\section{Flashing ratchet} 
\label{SEC6}

We have discussed macroscopic systems so far. In contrast, when the number of particles becomes small, like e.g. in molecular machines \cite{Chipot}, the environment itself continuously drive the system away out of equilibrium and the fluctuations it generates over the system are very relevant, so that the probability of observing an apparent violation of the second principle becomes significant. Such small machines will spend part of their time actually running in reverse, i.e. now and then it is possible to observe that these small molecular machines are able to generate work by taking heat from the environment, in agreement with Kuramoto's identity \cite{Carberry}. 

The generation of mechanical work out of thermal fluctuations requires a) either a time-dependent modulation or another form of energy flux, b) or some spatial asymmetry with a temperature gradient \cite{Feigel}. In ST, cases a) and b) correspnd to systems which are coupled to 
one (or several) heat bath(s) of constant temperature each respectively. The aim of this Section is to apply \eqref{variationaloscillating} in the description of stable operation of such machines in a particular class of problems in the case a). Sec. \ref{Brown} below is devoted to b).

In case a) $ \nabla T = 0 $ and particles diffuse in a potential $ \vartheta $, which is periodic both in space and time \cite{Sinytsin} \cite{Parrondo} \cite{Dolbeault}. The particle distribution satisfies a Fokker-Planck equation which is solved by a particle distribution $ f $ and where $ \textbf{W} = \nabla \vartheta $, just as in \cite{Casas}, and the relaxed state (if any) may oscillate in time. As for the boundary condition, the system remains confined at all times inside a region of the state space, just like in Sec. \ref{SEC2}. The equations of chemical kinetics acts as equations of motion and provide the constraints. Since $ \nabla T = 0 $ and both the volume and the chemical composition are fixed (particle diffusion is neglected in agreement with \eqref{slowdiffusion}), the free Helmholtz energy $\textit{F}$ too is constant. Since $ \nabla T = 0 $ and the slow diffusion approximation allow us to neglect the amount of entropy produced per unit time by heat transport and particle diffusion respectively, the minimization in \eqref{variationaloscillating} reduces to minimization of the path-ensemble-averaged, time-averaged total dissipated power. 

\begin{figure}[!h]
\centering
\includegraphics[scale=0.9]{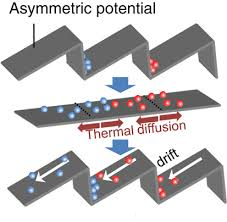}
\caption{\textit{In a flashing ratchet, $ T $ is both uniform and constant. The distribution function $ f $ provides information on the position of particles (the small spheres in the figure, which displays a purely 1D motion for simplicity) located inside a finite volume and subject to a potential $ \vartheta $. In turn, $ \vartheta $ is controlled by the external world, is periodic both in space and time, and is asymmetric in space. In the figure, time follows the blue arrows. At all times particles undergo both thermal diffusion and drift motion (under the effect of $ \nabla \vartheta $). At t = 0 (top) particles prefer to stay near the minima of $ \vartheta $ (red and blue particles are displayed for clarity near distinct minima of $ \vartheta $, and $ \vartheta $ is displayed as a grey, undulated surface). As time goes by (middle), the external world flattens $ \vartheta $, and particles undergo thermal diffusion only. After a period (bottom), the external world restore the potential $ \vartheta $ at t = 0; by now, however, particles may be shifted on average. The overplay of $ \vartheta $ and $ D $ plays a crucial role even if $ D $ is small. Maximum stability against thermal (hence diffusion-related) fluctuations corresponds to minimum averaged amount of power to be supplied by the external world in order to compensate losses - see text.}}
\label{flashingratchet}
\end{figure}

Indeed, it has been shown that the evolution of a particular kind ('flashing ratchet' see Fig.~\ref{flashingratchet}) of molecular motor - where the potential oscillates in time between zero and a periodic, asymmetric function of position - follows a path of '\textit{minimum dissipation of the total energy}' with the constraint of given $\textit{F}$ - see equation (4) of \cite{Dolbeault}. Admittedly, the treatment of Ref. \cite{Dolbeault} deals mainly with the 1D case, but in its own words '\textit{many of our results apply when $ \Omega $ is a bounded domain in $ \mathbb{R}^N , N \geq 1 $}'. The same principle has been applied to more complicated molecular motors - see equation (4.4) of \cite{Chipot}; on a qualitative base, generalization to a Fokker-Planck system with $ N = 3 $ chemical species seems possible - see Appendix \ref{QUAL}. 

Mathematically, the analysis of \cite{Dolbeault} relies on the results of \cite{Jordan}. Physically, $ \nabla T = 0$ in the flashing ratchet; moreover, the latter is compared to a overdamped mass-string-dashpot system where viscous dissipation only occurs; the crucial point is that '\textit{molecular motors move in a very viscous environment}' in the words of Sec. 2.2 of Ref. \cite{Dolbeault}, so that minimization of the dissipated power resembles Kortweg-Helmholtz' principle invoked in the $ \nabla T = 0$ DPF pinch - see Sec. \ref{SEC5}. 

Crucially, $ D $ never vanishes exactly, even if it is small; smallness of diffusion means just that we are focussing on slowly relaxing perturbations, as anticipated in Sec. \ref{SEC4}. Rather, the overall behaviour of the system turns out to be the outcome of the interplay between the tendency of particles tovards concentration at the minima of the potential and diffusion, which tends to spread and dissipate density. If $ T $ is uniform and constant, and if the ratchet keeps on oscillating in a stable way - so that the average amount of mechanical energy exchanged with the external world is also constant - then the same holds for $\textit{F}$ - this explains the constraint. Once the ratchet has started oscillating at a given period, physical intuition requires that the external world must do some work per period on the system in order to compensate the effect of thermal fluctuations, which tend to destroy the ordered oscillatory behaviour of the ratchet; ratchet stability against fluctuations is maximum when the required time-averaged amount of power is a minimum. 

A final remark. In ST, if a $ \nabla T = 0 $ system is in thermal contact with a heat reservoir at temperature $ T $, equations (3) and (7) of \cite{Williams} - together with the fact that $ \Pi $ is the volume integral of $ \sigma $ - show that $ \overline{ \widehat{\left[\Pi\right]}} $ is equal to the time-averaged amount of work done on the system by the external world divided by $T$. The same equations ensure also that $ \overline{ \widehat{\left[\Pi\right]}} > 0 $ even if the discussion of \cite{Williams} takes explicitly into account the possibility that $ \sigma $ may not be positive-definite.

\section{Brownian motor}
\label{Brown}

Our result \eqref{variationaloscillating} provides information on $ \nabla T \neq 0 $ systems too. Again, and in analogy with the discussion of macroscopic systems in Sec. \ref{SEC5bis} , if $ D \propto T $ then $ T $ affects \eqref{physics} through $ D $ only; then, in the slow diffusion approximation both $ T $ and $ \nabla T $ leave the validity of \eqref{variationaloscillating} unaffected. For example, Feynman describes \cite{Feynman} a ratchet coupled via an axle to a paddle wheel immersed in a fluid. Because of the Brownian motion of molecules, the paddle wheel has an equal
probability to be rotated to the right or the left. Owing to the presence of the pawl preventing one rotation direction, say the left, the ratchet will rotate on the right, seemingly violating the second law of thermodynamics. The apparent contradiction is
solved and equal rotation probability restored by including thermal fluctuations of the ratchet. However, if its temperature is smaller than that of the fluid, unidirectional rotation becomes possible, enabling conversion of heat into work. Rather than an applied, oscillating potential, it is therefore the non-uniformity of temperature which allows the motor to work. No $ \textbf{W} $ is required anymore; for simplicity we are going to take $ \textbf{W} = 0 $ altogether. The aim of this section is to apply \eqref{variationaloscillating} to the description of stable, relaxed states of such mesoscopic system, henceforth referred to as 'Brownian motor'. To this purpose, we follow the treatment of \cite{Feigel} and \cite{Meurs}.

In a system made by $ N_{p} $ parts, the $ i-$th part ($ i = 1,\ldots N_{p} $) being in thermal contact with a heat bath at given temperature $ T_{i} $, we are allowed to invoke \eqref{variationaloscillating} as far as the system remains confined at all times inside a region of the state space and $ f $ obeys a conservation equation like \eqref{physics} - see equations (3)-(4) of \cite{Feigel}; as for both the self-consistency of slow diffusion approximation and the physical meaning of $ f $, see below. Minimization in \eqref{variationaloscillating} implies minimization of the contribution of each part to $ \overline{\widehat{\left[\Pi\right]}} $ separately. In turn, this implies minimization of the path-ensemble-averaged, time-averaged total dissipated power in the $ i-$th part for all $ i $'s, because in the bulk of the $ i-$th part no entropy is produced by heat transport (as $ \nabla T_{i} = 0 $ across the $ i-$th part) and the entropy produced by particle diffusion is negligible in the slow diffusion approximation. Finally, and in analogy with our discussion of flashing ratchets above, minimization of the path-ensemble-averaged, time-averaged total dissipated power follows - as the latter is just the sum of the minimized amounts of power dissipated in each part. 

\begin{figure}[!h]
\centering
\includegraphics[scale=0.6]{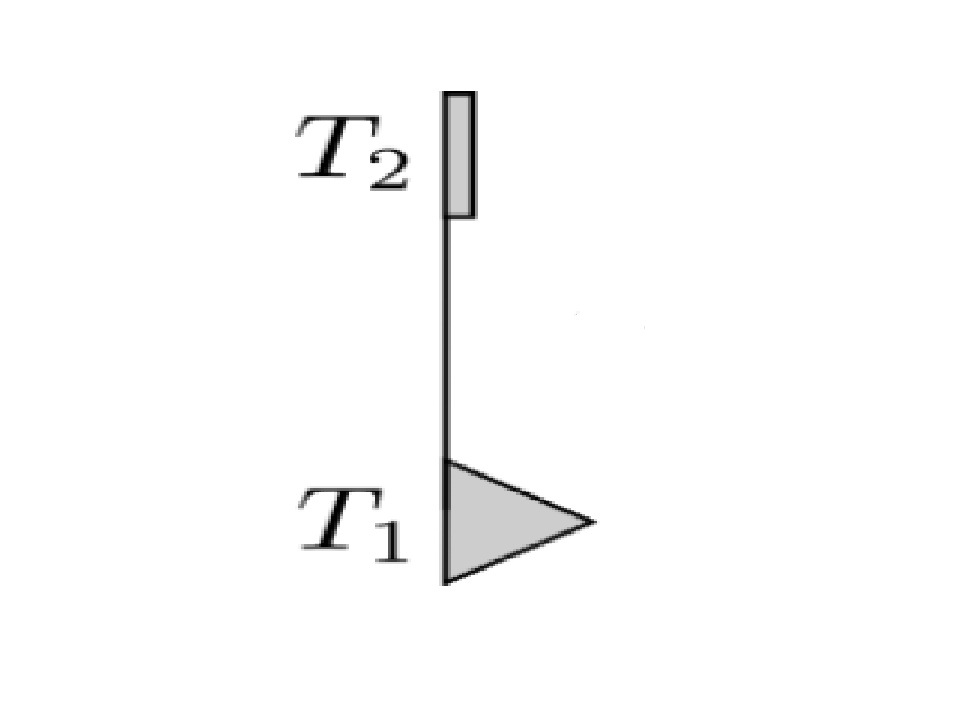}
\caption{\textit{A Brownian motor \cite{Feigel} made of $ N_{p} $ parts (here $ N_{p} = 2$ for simplicity). During the motion of the system (not displayed in the figure), it is assumed that the $ i-$th part ($ i = 1,\ldots N_{p} $) is in thermal contact at all times with a heat bath at temperature $ T_{i} $, and may undergo a viscous drag with a orientation-dependent viscous coefficient $ \gamma_{i} $. The values of the $ T_{i} $'s are given, and differ from each other. Moreover, the $ i-$th part has its own shape (here, a triangle and a rectangle are displayed for $ i = 1 $ and $ i = 2 $ respectively). Here, part 1 is free to change its own orientation; for example, the tip of triangle $ 1 $ may point either horizontally (as displayed in the figure) or vertically. On average, dissipation gets spontaneously minimized through changes in this orientation.}}
\label{BROWNMOTOR}
\end{figure}

To fix the ideas, let us suppose that the $ i-$th part is immersed in a fluid heat bath and undergoes a viscous drag with friction coefficient $ \gamma_{i} = \emph{b}_{i} \sqrt{T_{i}} > 0 $, where the $ i-$th coefficient $ \emph{b}_{i} $ of proportionality depends on the detailed geometry and orientation in space of the $ i-$th part - see Fig.~\ref{BROWNMOTOR}, as well as equation (12) of Ref. \cite{Feigel} and equation (34) of \cite{Meurs}, where the fluids are perfect gases. The $ i-$th part is free to change its own orientation in space during the operation of the motor. The amount of power dissipated in the $ i-$th part is an increasing function of $ \gamma_{i}$; then, minimization of the path-ensemble-averaged, time-averaged total dissipated power in each part - required by \eqref{variationaloscillating} - implies minimization of $ \gamma_{i} $ for all $i$'s. This minimization occurs as all the parts adjust their own orientation in space. This orientation is unambiguosly described by a set of Euler angles, which are the $ x^{i} $'s of the state space of the system, and $ f $, the probability distribution of orientation, depends on both the $ x^{i} $'s and time. If only 1 part out of $ N_{p} $ is free to change orientation then there are 3 Euler angles; the state space is 
$ \left[0,2\pi\right) \cup \left[0,2\pi\right) \cup \left[0,\pi\right] $ and can be linked with $ \mathbb{R}^3 $ by a diffeomorphism. It is useful to define the quantity $ T_{eff} \equiv \frac{\sum _{i} \gamma_{i} T_{i} }{\sum _{i} \gamma_{i}} $; it is an increasing function of fluctuation amplitude - see equation (2) of \cite{Feigel}. Since $ T_{i} \propto \gamma_{i}^2 $, minimization of $ \gamma_{i} $ implies minimization of $ T_{eff} $.

We have seen that minimization of $ T_{eff} $ is a necessary condition for the validity of \eqref{variationaloscillating} in our system. We show that is also a sufficient condition. Indeed, the facts that $ T_{i} \propto \gamma_{i}^2 $ and that $ \gamma_{i} > 0 $ make the minimization of $ T_{eff} $ to imply the minimization of viscous heating power. Since the latter is just the sum of the amounts of viscous heating power dissipated in each gas and each amount is non-negative, minimization of each amount separately follows. This minimization is achieved through changes in the detailed geometry of each part \cite{Feigel}, hence all $T_{i}$'s remain unaffected. As a consequence, minimization of the amount of viscous heating power in each part implies minimization of the amount of entropy produced in each part; after summing on all parts and averaging, \eqref{variationaloscillating} follows.

Indeed, it has been shown - in the particular case where all Euler angles but one are blocked at least - that: a) a molecular motor whose parts are in thermal contact with perfect gases acting as heat baths at different temperatures tends to minimize $ T_{eff} $ ; b) this minimization implies also the minimization of the diffusion coefficient, in agreement with the slow diffusion approximation; c) the relaxed state may violate  MinEP - see equations (5), (7) and the comment to equation (18) of \cite{Feigel}. Reults c) holds in analogy with the results of \cite{Qian2002} concerning Joule heating and referred to in Sec. \ref{SEC4} .  

\section{Conclusions}
\label{SEC7}

The second principle of thermodynamics provides information concerning a physical system at thermodynamic equilibrium; such information do not depend on the choice of coordinates with the system is described with. Far from thermodynamic equilibrium, the same holds for the description provided by Glansdorff and Prigogine's minimum entropy production principle (MinEP) for the stable, relaxed state of a macroscopic system which satisfies Onsager symmetry \cite{Gyarmati}. (Conservatively, here we stick to the familiar Gibb's entropy). To date, no proposed extremum property for far-from-equilibrium, stable relaxed states of systems which violate Onsager symmetry seems to enjoy such invariance. For example, a first-principle proof \cite{Martyushev3} of the so-called maximum entropy production principle (MEPP) \cite{Martyushev} seems to rely precisely on violation of this invariance. Physical intuition, in contrast, makes us to look for extremum properties - for stable, far-from-equilibrium relaxed states - which are invariant under change of coordinates. The price to be paid is likely to be the loss of generality. Indeed, the results of \cite{Di Vita2} suggest that variational principles concerning particular contributions to the entropy balance may hold for selected classes of problems.

Another kind of invariance - namely, invariance of microscopic physics under time reversal - plays a crucial role in Stochastic Thermodynamics (ST), which provides a framework for extending the notions of classical thermodynamics to the level of individual trajectories across the space state of a selected class of (usually mesoscopic) physical systems, namely those systems where a non-equilibrium process occurs which is coupled to one (or several) heat bath(s) of constant temperature. 
Remarkably, ST is often concerned with oscillating relaxed states, in contrast with MinEP and MEPP, which deal with steady relaxed states. Rather than to conditions of stability for stable relaxed states, ST leads to a large number of exact, experimentally confirmed results - usually referred to as 'fluctuation theorems' ('FTs') - starting from the invariance of the microscopic dynamics under time-reversal. FTs allow to quantify the probability of observing occasional 'reverse' behaviour, i.e. i.e. an entropy production opposite to that dictated by the second law of thermodynamics; it decreases exponentially as the system size increases \cite{EVANS} \cite{Seifert} \cite{VanderBroeck}.   

The success of FTs suggests to take advantage of the properties of invariance of a system while looking for some property of its stable, far-from-equilibrium relaxed states. In particular, it seems worthwhile to investigate the consequences of the invariance under diffeomorphisms involving the coordinates of the space state. Generally speaking, these diffeomorphisms can map macroscopic systems (where the volume of the system in the space state is large enough and fluctuations can be neglected in some cases at least) and mesoscopic systems (where the volume is small and fluctuations, possibly of quantum origin, cannot be neglected altogether) onto each other. Correspondingly, if such a property actually exists and is invariant under change of cooordinates then it provides information on the stability of both macroscopic and mesoscopic systems. Moreoverm given the fact that relaxed states of mesoscopic systems are often oscillating states, this - so far hypothetical - property should describe oscillating relaxed states - and steady states as a particular, zero-frequency case - in order to cope with both macroscopic and mesoscopic systems. A connection could be established between different problems of stability. Finally, of course, this property should agree with ST.

The aim of this paper is twofold: a) to write down this invariant property for the relaxed states of a selected class of physical systems, and b) to discuss some applications to stability. We limit ourselves to systems described by a PDE problem \eqref{physics} made of a simple Fokker-Planck equation and a boundary condition (the system remains confined at all times within a given region of the state space). It has been shown \cite{Casas} \cite{Polettini} \cite{Qian2002} that if a generic, differentiable transformation 
$\textit{M}$ of both the coordinates which the probability distribution function $ f $ depend upon and of the quantities involved in \eqref{physics} leaves both \eqref{physics} and time unchanged, then the amount $ \Pi $ (often referred to as $ \frac{d_{i}S}{dt} $ in the literature) of entropy produced per unit time by irreversible processes inside the bulk of the system is  invariant under $ \textit{M} $ . 

We have investigated the consequences of this invariance. To start with, each $\textit{M}$ is a relationship of equivalence, hence it establishes a partition of the set of the solutions of \eqref{physics} in equivalence classes; each class is unambiguously labelled by a value of $ \Pi $. Secondly, given the possibility of writing \eqref{physics} in dimensionless form (and of avoiding therefore problems with dimensional analysis), it follows that different elements of the same class of equivalence may represent different physical systems with the same value of $ \Pi $, as far as all of them satisfy \eqref{physics} even if the other details of the dynamics are different. 
For further discussion, firstly we discuss the necessary condition for the stability of a steady relaxed state of a macroscopic system; generalization follows.

If relaxation occurs towards some final, relaxed, steady state which is stable against small perturbations then unsteady, relaxing states are mapped onto unsteady, relaxing states and relaxed, stable states are relaxed onto stable, relaxed states. For example, thermodynamic equilibria are mapped onto thermodynamic equilibria ($ \Pi = 0 $). 
In particular, if at least one ('pivotal') of the systems which satisfy \eqref{physics} exists where constrained minimization of $\Pi$ (the constraints being provided by the equations of motion) is a necessary condition for the stability of relaxed states against small, slowly evolving perturbations, then the same property is shared by the stable, relaxed states of all physical systems which are obtained from the pivotal system via a map $ \textit{M} $; even if the detailed equations of motion may be different, the minimized quantity is the same. This corollary follows ultimately from the invariance of $ \Pi $, not from validity of Onsager symmetry nor from the detailed nature of the involved systems.

Further discussion takes advantage of the fact that we are looking for necessary conditions for stability, hence we are free to select the perturbation which stability is to be checked against. For simplicity, we assume that the perturbation is slow, i.e. that its typical time scale is longer than other typical time-scales of the dynamics of the system (for example, the oscillation period in a oscillating relaxed state). This is equivalent to assume that the diffusion term in the Fokker-Planck equation is relatively small (diffusion stands for irreversible phenomena which damp perturbations of stable states); this approximation is referred to as 'slow diffusion approximation' below.

Now, we show that the electrons in the pinch of a Dense Plasma Focus ('DPF') \cite{Soto} form a pivotal system. Depending on the relative relevance of electromagnetic and viscous force \cite{DiVitaHartmann}, the spatial structure of the system may either filamentary or plasmoid-like \cite{Di Vita}. Stable, steady relaxed configurations minimize $\Pi$. (Experiments \cite{Herold} \cite{Jakubowski} suggest that these structures are indeed exceptionally stable against resistive decay as predicted by magnetohydrodynamics ('MHD'), so that the words 'stable' and 'steady' seem to be justified). 

Generalization from steady to oscillating relaxed states of macroscopic systems requires replacement of $\Pi$ with its time-average $ \widehat{\left[\Pi\right]} $, the time-average being taken on a time scale which is both much longer than one period of oscillation of the relaxed state and much shorter than the typical relaxation time scale of the perturbation \cite{Tykodi}. Analogously, generalization to mesoscopic systems requires further replacement of $ \widehat{\left[\Pi\right]} $ with its path-ensemble average \cite{Crooks} $ \overline{\widehat{\left[\Pi\right]}} $ . These generalizations rely on the fact that the map commute with both time-averaging and path-ensemble-averaging, near a stable relaxed state at least. When it comes to mesoscopic systems, we could say that the equivalence classes referred to above act as "universality classes" which these systems can be grouped in.

A striking consequence of the generalisation to mesoscopic systems is that the problems of stability of macroscopic and mesoscopic systems may be dual to each other, i.e. share a common solution. The problem of stability against slow perturbations of the relaxed state of a mesoscopic system described by \eqref{physics} which remains confined at all times within a given region of the state space (e.g., a Brownian motor) and the problem of stability of the relaxed state of a weakly dissipative fluid with no net mass source, no mass flow outside the boundary and negligible particle diffusion (e.g., the subphotospheric, convection-ruled, dynamo-affected layer of a G-class rotating star) are dual to each other. Even if the star produces an immensely larger amount of entropy per unit time, it is not the actual, averaged value of $ \Pi $ which is relevant to the problem of stability, but its minimum property. (Analogously, it is the maximum property of entropy $ S $, not its actual value, which is relevant to the stability of thermodynamic equilibrium). Qualitative arguments even suggest that constrained minimization of $ \overline{\widehat{\left[\Pi\right]}} $ are indeed in agreement with ST, even if no pivotal system is available - see Appendix \ref{QUAL} . This minimization is therefore the looked-for invariant property of stable relaxed states. Fig.~\ref{provaschema} displays a general overview of the results of our discussion. In a nutshell, invariance bridges the gap between the problems of stability of some selected macroscopic and mesoscopic systems. 

In order to discuss applications, we start with the following remark. As far as we investigate their stability against slowly evolving perturbations, a macroscopic, classical fluid - with Joule and viscous heating, negligible mass diffusion, no net mass source and negligible mass flow across the boundary - and a mesoscopic system satisfying \eqref{physics} - with $ \mathbb{R}^3 $ as the state space - enjoy the same property: both can be mapped onto a pivotal system. In the macroscopic fluid, equation \eqref{physics} reduces to a simple mass balance; in the mesoscopic system, investigation of stability against slow perturbations allow us to invoke the slow diffusion approximation. Consequently, they share also the same necessary condition of constrained minimization of $ \overline{\widehat{\left[\Pi\right]}} $ for the stability of relaxed states. The equations of motion provide the constraints, which depend therefore on the particular system of interest. It does not matter what the detailed dynamics is like; what is relevant here is that such mapping is possible. Examples of the macroscopic fluids quoted  above are the filamentary configuration of the electrons in the DPF pinch and a (possibly unmagnetized) fluid near the onset of Rayleigh-Bénard convection; indeed, it is shown in  Appendix \ref{CGLEapp} that the problems of stability of the two systems are dual to ech other, even if the dynamics is utterly different.

Beyond thermodynamic equilibrium (where of course $ \overline{\widehat{\left[\Pi\right]}} \equiv 0 $), in the following we are going to discuss four particular problems as particular cases of the macroscopic and the mesoscopic systems hinted at above. In each case, we have derived from the constrained minimization of $ \overline{\widehat{\left[\Pi\right]}} $ a necessary condition of stability, not necessarily in the form of a variational principle. The criterion of stability we obtain depends on the particular problem. Noteworthy, both MinEP and MEPP apply in no case. In particular, according to the language of Ref. \cite{MS} all macroscopic systems we deal with in the following are 'compound' systems, where MEPP fails. This is not surprising, as both MinEP and MEPP are concerned with the entropy produced per unit time by all irreversible processes in the system, while $ \Pi $ is the amount produced per unit time by the irreversible processes only which occur in the bulk. In all cases we retrieve results previously available in the literature, independently derived from either observations or dedicated proofs from first principles.  

\begin{table}[!h]
\caption{\textit{Classification of some necessary criteria of stability of relaxed states against slow perturbations for systems which obey the equation \eqref{physics} $ \frac{\partial f}{\partial t} + \nabla \cdot \textbf{J} = 0 $ with $ \textbf{n} \cdot \textbf{J} = 0 $ at the boundary.  
Here $ f $, $ \textbf{J} = \textbf{W} f - D \nabla f$ and $ \textbf{n} $ are a distribution function, a probability current density and the unit normal vector perpendicular to the boundary of the region $ \Omega \subset \mathbb{R}^3 $ (with measure $ V $) where the system is confined; moreover, $ D $ and $ \textbf{W} $ are a diffusion coefficient and a vector field respectively. 
We denote by $ \widehat{\left[\emph{y}\right]} $ and $ \overline{\textit{y}} $ the time average and the path-ensemble-average \cite{Crooks} respectively of the generic quantity $ \emph{y} $. Following \cite{Casas}, we introduce the amount 
$ \Pi = \int \frac{k_{B}}{D} \frac{\vert \textbf{J} \vert^2}{f} \mbox{d}V $ 
of the Boltzmann-Gibbs entropy $ S $ of the system produced in the bulk of the system, where $ k_{B} $ is Boltzmann's constant. 
All necessary criteria are retrieved as particular examples of the constrained minimization of $ \overline{\widehat{\left[\Pi\right]}} $, the constraints being provided by the equations of motion in each case.  
Stable, relaxed states of macroscopic (mesoscopic) systems correspond to 
$ \overline{\widehat{\left[\Pi\right]}} - \widehat{\left[\Pi\right]} = 0 
\left( \neq 0 \right) $. 
Stable, relaxed, steady (oscillating) states correspond to $ \widehat{\left[\Pi\right]} - \Pi = 0 \left( \neq 0 \right) $.
Macroscopic systems of interest have no net mass source, no mass flow across the boundary and negligible mass diffusion, provided that $ f $ and $ \textbf{J} $ are mass density and flow respectively.}}
\bigskip
\begin{tabular}{|l|c|c|c|c|}
\hline
Quantity	&	
$ \left( A \right) $ \cite{Di Vita} \cite{Taylor00} \cite{Comisso} &	
$ \left( B \right) $ \cite{Chandrasekhar} \cite{Jordan000} & 
$ \left( C \right) $ \cite{Dolbeault} & 
$ \left( D \right) $ \cite{Feigel} \\
\hline
$ 
\vert \nabla T \vert $	&	
$\qquad = 0 \qquad $	&	$ \qquad \neq 0 \qquad $	&	$ \qquad = 0 \qquad $	&	$\qquad \neq 0 $\\
$ 
\vert \nabla \wedge \textbf{W} \vert$	&	
$\qquad \neq 0 \qquad$	&	$ \qquad \neq 0 \qquad$	&	$\qquad = 0 \qquad$	&	$\qquad = 0$\\
$ 
\overline{\widehat{\left[\Pi\right]}} - \widehat{\left[\Pi\right]} $ 	&	
$\qquad = 0 \qquad$	&	$\qquad = 0 \qquad	$	&	$\qquad \neq 0 \qquad $	&	$\qquad \neq 0$\\
$ 
\widehat{\left[\Pi\right]} - \Pi $	&	
$\qquad = 0 \qquad$	&	$\qquad \neq 0 \qquad$    &	$ \qquad \neq 0 \qquad $ &	$\qquad \neq 0$\\
\hline
\end{tabular}
\label{tabellasinottica}
\end{table}

Results are collected in Table \ref{tabellasinottica} . The list is
not - and is not meant to be - complete. Each column of Table \ref{tabellasinottica} refers to a different problem far from thermodynamic equilibrium, where each of the quantities on the left may either vanish ('$ = 0 $') or differ from zero ('$ \neq 0 $').  For each problem, results are discussed below; we start from the 
minimization of $ \overline{\widehat{\left[\Pi\right]}} $ and 
retrieve a corresponding necessary criterion of stability, which finds its confirmation in the Refs. referred to in the first line of the Table. For comparison, $ S = \max $ is the necessary criterion of stability of thermodynamic equilibrium ($ \dot{S} \equiv 0, \Pi \equiv 0$).

(A) Electrons in a DPF pinch \cite{Soto}. Rather than the identically vanishing $ \nabla T $, it is 
$ \nabla \wedge \textbf{W} \neq 0 $ which keeps the stable, steady relaxed state far from thermodynamic equilibrium in this macroscopic system. Here $ \textbf{W} $ is a suitably defined linear combination of velocity and vector potential and $ f \propto $ mass density. ('Steady' means here 'with lifetime much longer than predicted by MHD'). 
The necessary criterion of stability of the relaxed state is $ \int P_{h} \mbox{d}V = \min. $ with fixed $ T_{\mbox{boundary}} $ \cite{Di Vita}, where 
$ P_{h} $ is the sum of viscous heating power and Joule heating power density, and 
$ T_{\mbox{boundary}} $ is the value of the temperature $ T $ at the boundary. If viscous heating is (not) negligible in comparison with electromagnetic force, then a plasmoid-like configuration (a filamentary structure) arises. In the plasmoid case \cite{Jakubowski}, th enecessary criterion of stability a monotonically increasing function of resistivity achieves a minimum - in agreement with the results of \cite{Comisso}; moreover, Taylor's principle of minimum magnetic energy for given minimum helicity is approximately satisfied \cite{Taylor00}. In the filamentary case  \cite{Herold}, current filaments are similar to those of a superconductor \cite{Froehlich} ; the relaxed state belongs to the same class of equivalence of a (possibly unmagnetized) fluid near the onset of Rayleigh-Bénard convection, in spite of the different equations of motion and of the different values of $ \nabla T $.

(B) Rayleigh-Bénard convection in a rotating shell of magnetized fluid affected by dynamo. The subphotospheric, convection-ruled zone of a rotating star \cite{Priest} provides an example. Here $ \textbf{W} $ and $ f $ are $ \propto $ the velocity and the mass density repsectively. Both $ \nabla T \neq 0 $ and $ \nabla \wedge \textbf{W} \neq 0 $ keep the oscillating relaxed state of this macroscopic system far from thermodynamic equilibrium; rotation of both the shell and the convective cells rule the oscillation frequencies. It turns out that the absolute value of $ \nabla T $ is minimized, with the constraint that the time-averaged dissipated power is equal to the time-averaged mechanical power provided by buoyancy; this result had been previously found in \cite{Di Vita2} under the more restrictive assumption of local thermodynamic equilibrium \cite{Glansdorff}, and is in agreement with \cite{Chandrasekhar}. As a consequence, the averaged magnetic field is inversely proportional to the Rossby number, in agreement with observations for F and G/K dwarf stars \cite{Jordan000} \cite{Reiners} .

(C) Flashing ratchet. An externally applied, oscillating potential keeps this mesoscopic system far from thermodynamic equilibrium, even if $ \nabla T = 0 $ and $ \nabla \wedge \textbf{W} = 0 $ ($ \textbf{W} = \nabla \vartheta $, where $ \vartheta $ is an externally applied potential, which is periodic both in time and in space, and is asymmetric in space; $ f $ is a particle distribution). In order to sustain regular ratchet oscillations, the external world must do some work per period on the system in order to compensate the effect of thermal fluctuations, which tend to destroy the ordered oscillatory behaviour of the ratchet; ratchet stability against fluctuations is maximum when the required time-averaged amount of power is a minimum. It turns out  that stable relaxed states minimize the dissipated power with the constraint of given Helmholz' free energy. This is in agreement with the results of Ref. \cite{Dolbeault}.

(D) Brownian motor. This mesoscopic system is made of $ N_{p} $ parts ($ i = 1,\ldots N_{p} $), where the $ i $-th part is  immersed in a fluid heat bath at temperature $ T_{i} $ and viscous coefficient $ \gamma_{i} $. Rather than an applied, oscillating potential, it is therefore the non-uniformity of temperature which allows the motor to work, by enabling conversion of heat into work. The differences between the constant $ T_{i} $'s
keep the system far from thermodynamic equilibrium. Stable relaxed, oscillating states minimize a a quantity (the 'effective temperature' $ T_{eff} $) - which has the dimension of a temperature and which depends on the $ \gamma_{i} $'s - under the constraint of given $ T_{i} $'s. This is in agreement with the results of Ref. \cite{Feigel}, where $ \textbf{W} = 0 $ and $ f $ refers to the probability of orientations in space; this treatment stems from the seminal analysis of Feynman \cite{Feynman}.

We have shown that invariance against changes of coordinates plays a crucial role when it comes to understand many different criteria for stability for dissipative macroscopic and mesoscopic systems. In a nutshell, invariance allows us to see that the problem of the stability of the far-from-thermodynamic-equilibrium relaxed state of a physical system may be dual to the corresponding problem in another system, even if the two systems may differ considerably. Then, once the problem has been solved for one system, the solutions of the problems for a whole class of physical system is at hand. Here we have discussed invariance of suitably averaged $ \Pi $ for a wide class of systems. There is no fundamental reason to think that invariance of other quantities for other classes of systems is impossible, and that it leads to no further criteria for stability. 

Finally, it is clear, by now, that our pivotal system acts just as a Trojan horse which allows us to smuggle the stability criterion \eqref{variationaloscillating} into the world of the relaxed solutions of \eqref{slowlyvaryingf} under the slow diffusion approximation. DPF pinch electrons (column (A) in Table Table \ref{tabellasinottica}) are by no means special: once the existence of the pivotal system is proven, its detailed dynamics is not relevant, as expected in a thermodynamic approach. We could as well start e.g. from the systems described in columns (B, (C) or (D), where the stability criteria have been speparately and independently established. It is even possible that no 'pivotal' system is really required - as suggested by the qualitative discussion of Appendix \ref{QUAL} - and that a deeper connection exists between invariance and stability exists. Thus, further extension of the method outlined here to other problems is conceivable, and will be the matter of future work.

\section*{Acknowledgments}

Fruitful discussions with Prof. M. Polettini (Univ. Luxembourg) and Dr. E. Cosatto (Ansaldo Energia, Genova) are warmly acknowledged.

\appendix 

\section{More about MEPP}
\label{MEPPapp}

MEPP meets no general consensus yet. For example, Ziegler's version \cite{Ziegler} of MEPP has '\emph{its statistical substantiation only if the deviation from equilibrium is small}' \cite{Martyushev} and is criticised in \cite{Polettini00}. According to Ref. \cite{MS}, Ziegler's MEPP can be considered just as a working hypothesis for nonlinear nonequilibrium thermodynamics. Moreover, validity of MEPP is questionable in the so called 'compound processes'. By definition, the entropy production of a compound process is a sum of functions dependent only on some (but not on all) of the thermodynamic fluxes, and the fluxes which one of these functions depends upon do not affect the other functions. Simultaneous chemical and thermal processes can be the simplest example of compound processes: the former are scalar and the latter are vector; and according to Curie's principle, these processes cannot influence each other. Finally, the MEPP-supporting results of \cite{Dewar}, which rely on the so-called 'information thermodynamics' approach \cite{Jaynes00}, are criticized in  \cite{Grinstein}. In particular, it has been observed \cite{Di Vita2} that information thermodynamics provides the fundamental tenet of equiprobability of microstates with no physical basis similar to what is provided e.g. by Liouville's theorem in equilibrium thermodynamics; in the harsh words of \cite{Polettini0}, the author of \cite{Jaynes00} '\emph{swept the dirt under the carpet}'. 

According to a further, 
'steepest entropy ascent' 
approach \cite{Beretta1} \cite{Beretta2}, irreversibility in Nature is a fundamental microscopic dynamical feature and as such it must be built into the fundamental laws of time evolution. 
In this approach, MEPP ultimately arises from a fundamental non-unitary extension of standard Schroedinger unitary dynamics. 
In the words of \cite{Beretta1}, 
'\emph{every trajectory unfolds along a path of steepest entropy ascent compatible with the constraints} 
[...]. 
\emph{For an isolated system, the constraints represent constants of the motion.}
[...] 
\emph{the qualifying and unifying feature of this dynamical principle is the direction of maximal 
entropy increase.}
[...]
\emph{The challenge with this approach 
is to ascertain if the intrinsic irreversibility it implies at the single particle 
(local, microscopic) level is experimentally verifiable, or else its mathematics must
only be considered yet another phenomenological tool}'. Recent research aims precisely at comparison of predictions based on steepest descent approach and experiments where relaxation is postulated to follow a particular law - we refer to \cite{VonSpakowski} and in particular to its equation (12). 
When it comes to classical physics, the constraints acting on the relaxation identify a hyper-surface in the phase space of the system. In the words of \cite{Beretta2}, 
'\emph{on this surface we can identify contour curves of constant entropy, generated by intersecting it with the constant-entropy surfaces. Every trajectory} 
[...] \emph{lies}' 
on this surface 
'\emph{and is at each point orthogonal to the constant entropy contour passing through that point. In this sense, the trajectory follows a path of steepest entropy ascent compatible with the 
constraints}'. 
However, this conclusion relies on the postulate that relaxation obeys a particular law -see equation (25) of \cite{Beretta2}, retrieved also in equation (12) of \cite{VonSpakowski}.

Admittedly, general arguments supporting MEPP are discussed in \cite{Sawada}. Remarkably, however, this work seems to identify a property of $ \Phi $, rather than of 
$ \dot{S} $. In fact, equation (7) of \cite{Sawada} suggests maximisation of the amount per unit time of the entropy of the Universe due to exchange between the relaxed system and the external world - in our language, maximisation of $ -\Phi $ rather than of $ \dot{S} $. The same limitation seems to apply to many alleged applications of MEPP, including metallurgy \cite{Kirkaldy}, crystallization \cite{Hill} and crystal growth \cite{Martyushev2}, solidification \cite{Sekhar}, physics of the atmosphere \cite{Kleidon1} \cite{Kleidon2} \cite{Lorenz} \cite{Virgo} and plasma physics \cite{Yoshida}. 

For example, the growth of a crystal is obviously affected by what happens at its boundaries. Correspondingly, the selection criterion discussed in \cite{Kirkaldy} is related to the '\emph{conjecture that each lamella must grow in a direction which is perpendicular to the solidification front}'. Analogously, the results of \cite{Hill} are related to the growth velocity via the '\emph{rate of entropy production per unit area}', which is a surface quantity like $ \Phi $ and unlike $ \Pi $. Finally, in \cite{Martyushev2} a crystal grows and changes its surface so as to select a flow which maximises the entropy production, $ -\Phi $ being the only contribution to the entropy balance which depends on both a flow and the surface of the system. As for solidification, the relevant quantity is the '\emph{entropy generation term which depends on the solid-liquid region}' \cite{Sekhar}, i.e. on an interfacial boundary. As for the atmosphere, the only contribution to the entropy balance investigated in \cite{Kleidon1}, \cite{Kleidon2}, \cite{Lorenz} and \cite{Virgo} is due to energy fluxes across boundaries - see equations (5) and (14) of \cite{Kleidon1}, equations (1) and (2) of \cite{Kleidon2}, Fig. 1 of \cite{Lorenz} and equation (1) of \cite{Virgo} - and is therefore to be identified with $ -\Phi $; in contrast, the agreement of MEPP with models of a stratified atmosphere including viscosity (so that $ \Pi $ is not negligible) seems to be poor \cite{Marston}. As for plasmas, what is actually maximized in equation (7) of \cite{Yoshida} is the inhomogeneity of temperature in a boundary layer, i.e. a quantity related to $ \Phi $.

In contrast with MEPP, different extremum properties for $ \Pi $ and $ \Phi $ seem to hold for relaxed states of different physical systems under different boundary conditions -see e.g. Sec. 3 of \cite{Lucia} for a particular example and \cite{Di Vita2} for a more general discussion, which includes some of the results quoted above. For instance, the latter discussion involves also an extremum property similar to what is proposed in \cite{Sawada}, but concerning $ -\Phi $ rather than $ \dot{S} $; such property is suggested in \cite{Rebhan} for a shock wave in the limit of large Mach number, where the volume integral $ \Pi $ is as vanishingly small as the volume occupied by the shock front.

\section{More about \eqref{bilanciodientropia} and \eqref{physics}}
\label{Proofs2}

We follow the treatment of \cite{Casas} and \cite{Polettini}. The results listed below have been derived in \cite{Polettini} in the framework of the research concerning the models \cite{Andrieux1} \cite{Andrieux2} based on Schnakenberg's treatment of master equation \cite{Schnakenberg}. Equations \eqref{entropia}, \eqref{bilanciodientropia} and \eqref{physics} give:

\begin{equation} \label{internal0}
\dot{S} 
= 
\int \dfrac{k_{B}}{D} \dfrac{\vert \textbf{J} \vert^2}{f} \mbox{d}V
- \int \dfrac{k_{B} \left( \textbf{J} \cdot \textbf{W} \right)}{D} \mbox{d}V
\end{equation}

after integration by parts - for details, see the proof of equation (10) in \cite{Casas} for a Boltzmann-Gibbs entropy. The first term on the R.H.S. of \eqref{internal0} is the only term which is $ \geq 0 $ in all cases, because $ D > 0 $. Then, comparison of \eqref{bilanciodientropia} and \eqref{internal0} leads to \eqref{Pi} and \eqref{Phi}. 

According to \cite{Casas} and \cite{Polettini}, \eqref{physics} is invariant whenever 
\eqref{trasformazionediD}, \eqref{densitascalare} and \eqref{densitavettoriale} hold. 
Indeed, $ D $ is a scalar, then \eqref{trasformazionediD} holds. Moreover, the definition of $ \Lambda $ allows us to write:

\begin{equation} \label{elementodivolume}
\mbox{d}V ^{'} = \Lambda \mbox{d}V
\end{equation}

Now, invariance of the normalization condition $ \int f \mbox{d}V = 1 $ and equation \eqref{elementodivolume} imply \eqref{densitascalare}, i.e. $ f $ transforms like a scalar density. Finally, invariance of $ \dfrac{\partial f}{\partial t} + \nabla \cdot \textbf{J} = 0 $ and equation \eqref{densitascalare} imply \eqref{densitavettoriale}, i.e. $ \textbf{J} $ transforms like a vector density.

In turn, equations \eqref{trasformazionediD}, \eqref{densitascalare} and  \eqref{densitavettoriale} have a number of consequences. 

Firstly, invariance of our boundary condition follows from \eqref{densitavettoriale}.

Secondly, \eqref{entropia} and \eqref{densitascalare} imply $ S^{'} = S + <\ln \left( \Lambda \right)> $, i.e. $ S $ is not invariant -see equation (8) of \cite{Polettini}. 

Thirdly, \eqref{entropia} and Reynolds' transport theorem \cite{Smirnov} give 
$ \dot{S} = - k_{B} 
\int \left( \frac{\partial f}{\partial t} \right) 
\left[ 1 + \ln \left( f \right) \right] 
\mbox{d}V $. Together with \eqref{physics}, integration by parts and Gauss' theorem of divergence lead to 
$ \dot{S} = k_{B} 
\int \textbf{J} \cdot \nabla \ln \left( \frac{1}{f} \right)
\mbox{d}V $. In turn, the latter result, together with \eqref{elementodivolume} and \eqref{densitavettoriale}, leads to: 

\begin{equation} \label{dSdt}
\left( \dot{S} \right)^{'} = \dot{S} + 
k_{B} \int \textbf{J} \cdot \nabla \ln \left( \Lambda \right)
\mbox{d}V
\end{equation}

i.e. $ \dot{S} $ is not invariant - see equation (11) of \cite{Polettini}.

Fourth, \eqref{physics}, \eqref{trasformazionediD}, \eqref{densitascalare}, \eqref{densitavettoriale} and \eqref{dSdt} lead to \eqref{trasformazionediA} - see equation (10) of \cite{Polettini}.

Finally, after cumbersome algebra equations \eqref{elementodivolume}, \eqref{physics}, \eqref{densitavettoriale}, \eqref{dSdt} and \eqref{trasformazionediA} lead to \eqref{invariante2}. This result can also be obtained straightforwardly by observing that both $ D $ and the quantity $ \frac{\vert \textbf{J} \vert^2}{f} \mbox{d}V $ in \eqref{Pi} are invariant according to \eqref{trasformazionediD}, \eqref{densitascalare}, \eqref{densitavettoriale} and \eqref{elementodivolume}. In contrast with $ \Pi $, the lack of invariance of $ \dot{S} $, together with equations \eqref{bilanciodientropia} and \eqref{invariante2}, show that $ \Phi $ too lacks invariance.

\section{More about \eqref{invariante2}}
\label{CONF1app}

We can rewrite \eqref{invariante2} in a sligthly different way for systems which exchange only heat with a thermal bath at temperature $ T $. In such systems, $ \Phi $ is equal to the amount of heat flowing per unit time from the system towards the bath divided by $ T $ and is therefore equal to the time derivative $ \Sigma_{\mbox{b}}$ of bath entropy, provided that - as usual - we neglect all irreversible processes occurring inside the bulk of the bath, i.e. we assume $ \Pi = 0$ in the bath. Then, $ \Phi = \Sigma_{\mbox{b}} $, so that \eqref{bilanciodientropia} and \eqref{invariante2} give $ \left( \dot{S} + \Sigma_{\mbox{b}} \right)^{'} = 
\left( \dot{S} + \Sigma_{\mbox{b}} \right)$. Integration on time from $ t = 0 $ to a time $t$ gives:

\begin{equation} \label{invariante3}
\left(
S \left( t \right) - S \left( 0 \right) + 
\int_{0}^{t} \Sigma_{\mbox{b}} d\mbox{t}
\right)
^{'}
= 
S \left( t \right) - S \left( 0 \right) + 
\int_{0}^{t} \Sigma_{\mbox{b}} d\mbox{t}
\end{equation}

Just like \eqref{bilanciodientropia} and \eqref{invariante2} which is based upon, \eqref{invariante3} holds for both steady states and unsteady states. As for steady states, $ \dot{S} = 0$; moreover, we denote with $ \Sigma_{\mbox{ss}} $ the value taken by $\Sigma_{\mbox{b}} $. As for unsteady states, $\dot{S}$ may differ from $ 0 $ and we define $ \Sigma_{\mbox{ex}} \equiv \Sigma_{\mbox{b}} - \Sigma_{\mbox{ss}}$.  Finally, we write \eqref{invariante3} for both steady and unsteady state, subtract term by term and obtain 

\begin{equation} \label{invariante4}
\left( 
S \left( t \right) - S \left( 0 \right) + 
\int_{0}^{t} \Sigma_{\mbox{ex}} d\mbox{t}
\right)
^{'} 
=
\left( 
S \left( t \right) - S \left( 0 \right) + 
\int_{0}^{t} \Sigma_{\mbox{ex}} d\mbox{t}
\right)  
\end{equation}

Now, equation \eqref{invariante4} is just equation (11) of \cite{invarianceofsteady} (but for a slightly different definition of $ S $); the latter relationship deals with the invariance properties of the entropy balance in systems undergoing Markovian evolution according to an equation which generalises \eqref{physics} under transformations which include also rescaling of time-scales. 

\section{More about \eqref{variationaloscillating}}
\label{QUAL}

After multiplying $ \hat{\Pi} $ by $ -\frac{t}{k_{B}} $ and exponentiating, the reationship \eqref{invariant00} implies invariance of the dimensionless quantity 
$ \textit{X} \equiv \exp \left( - \textit{Y} \right) $ 
where 
$ \textit{Y} \equiv \frac{\hat{\Pi} t}{k_{B}} $. In other words, $ \textit{X} $ does not depend on the Jacobian $ \Lambda $ defined in Sec. \ref{SEC2}. 
After deriving $ \textit{X} $ once and twice on $ \Lambda $, we obtain therefore:

\begin{equation}
\label{auxil1}
0 = \frac{d \textit{X}}{d \Lambda} = 
- \textit{X} \frac{d \textit{Y}}{d \Lambda}
\end{equation}

and 
$ 0 = \frac{d^2 \textit{X}}{d \Lambda ^2} =
- \textit{X}
\frac{d^2 \textit{Y} }{d \Lambda ^2} +
\textit{X}
\left[
\frac{d \textit{Y}}{d \Lambda}
\right]^2 
\geq 
- \textit{X}
\frac{d^2 \textit{Y} }{d \Lambda ^2}
$ 
respectively; the latter relationship gives:

\begin{equation}
\label{auxil2}
\textit{X} \frac{d^2 \textit{Y} }{d \Lambda ^2} \geq 0
\end{equation}

After taking the path ensemble average of both sides of \eqref{auxil1} and \eqref{auxil2} 
we obtain 

\begin{equation}
\label{auxil3}
\overline{\textit{X} \frac{d \textit{Y}}{d \Lambda}} = 0 \quad ; \quad 
\overline{\textit{X} \frac{d^2 \textit{Y} }{d \Lambda ^2}} \geq 0
\end{equation}

Until now, we have assumed $ \Lambda $ to be constant. Nothing essential changes, however, if we let it to oscillate mildly and slowly in time, say 
$ \Lambda = 1 + \varepsilon \lambda \left( t \right), \lambda \left( t \right) \propto \cos \left( \frac{2 \cdot \pi \cdot t}{\tau_{\lambda}} \right), 0 < \varepsilon \ll 1, \tau_{aux } \ll \tau_{\lambda} \ll \tau$ - think e.g. of a self-similar rescaling with slow, alternating expansion and compression. A relaxed, stable state will be also oscillating with period $ \tau_{\lambda} $, at least as far as the external world keeps on driving such slow oscillations.

If no dissipation occurs inside the system, then $ \textit{Y} \equiv 0 $ and $ \textit{X} \equiv 1$ at all times. If dissipation occurs, our assumption of constant and uniform $ T $ leads to $ \Delta W_{diss} = T \cdot t \cdot \hat{\Pi} $, where $ \Delta W_{diss} $ is the amount of dissipated work from $ t = 0 $ to the time $ t $. 
It follows that $ \textit{Y} = \frac{\Delta W_{diss}}{k_{B} T}$ and 
$ \textit{X} = \exp \left( - \frac{\Delta W_{diss}}{k_{B} T}\right)$. 
Let us denote with $ \textit{y}_{R} $ 
and $ \left[ \overline{\textit{y}} \right]_{R}$ 
the quantities obtained by $ \textit{y} \left( \textit{C} \right)$ and 
$ \overline{\textit{y}} $ via time reversal respectively. Now, we may invoke the identity 

\begin{equation}
\label{auxil4}
\overline{\textit{X} \textit{y}} = \left[ \overline{\textit{y}_{R}} \right]_{R} 
\end{equation}

-see equation (15) of \cite{Crooks}- provided that some assumptions are made; this point is discussed below.  If $ \textit{y} = 1 $ then \eqref{auxil4} reduces to Jarzynski's equality 
$ \overline{\textit{X}} = 1 $.
Together, \eqref{auxil3} and \eqref{auxil4} lead to:

\begin{equation}
\label{auxil5}
\left[ \overline{ \left( \frac{d \textit{Y}}{d \Lambda} \right)_{R} } \right]_{R} = 0 
\quad ; \quad 
\left[ \overline{\left( \frac{d^2 \textit{Y} }{d \Lambda ^2} \right)_{R}} \right]_{R} \geq 0
\end{equation}

Since our stable relaxed state is permanently oscillating, however, we may drop the subscript $ _{R} $ altogether in \eqref{auxil5}, as time reversal leaves oscillating quantitites unaffected but for a change of phase. Accordingly, \eqref{auxil5} leads to:

\begin{equation}
\label{auxil6}
\overline{\frac{d \textit{Y}}{d \Lambda}} = 0 \quad ; \quad 
\overline{\frac{d^2 \textit{Y} }{d \Lambda ^2}} \geq 0
\end{equation}

Now, we invoke the argument of  Sec. \ref{SEC6ter}. The small perturbation $ \Lambda \rightarrow \Lambda + d \Lambda$ of a stable, relaxed state leaves the $ \textit{p} \left( \textit{C} \right) $'s remain unaffected, and we are allowed to write 
$\overline{\frac{d \textit{y}}{d \Lambda}} = \frac{d \overline{\textit{y}} }{d \Lambda}$  all along the perturbation. 
Accordingly, for a stable, relaxed state relationships \eqref{auxil6} and the definition of $ \textit{Y} $ lead to:

\begin{equation}
\label{auxil7}
\frac{d \overline{\hat{\Pi}}}{d \Lambda} = 0 \quad ; \quad \frac{d^2 \overline{\hat{\Pi}}}{d \Lambda ^2} \geq 0 
\end{equation}

i.e., $ \overline{\hat{\Pi}} $ is a minimum in stable relaxed states with respect to a small perturbation of $ \Lambda $. The assumption of oscillating $ \Lambda $ is relaxed  in the $ \tau_{\lambda} \rightarrow \infty $ limit (i.e. $ \Lambda $ is constant). It is relaxed too, provided that we write the generic, slowly evolving $ \Lambda \left( t \right) $ as a sum of harmonic functions of time with different periods, provided that each term is so small that we may consider the response of $ \overline{\hat{\Pi}} $ to $ \Lambda \left( t \right) $ as a sum of responses to each term.

Admittedly, our argument relies on an identity, \eqref{auxil4}, whose validity is limited by a number of additional assumptions \cite{Crooks}: a) the energy of the system is finite; b) the dynamics are Markovian; c) if the system is already in equilibrium and is unperturbed, then it must remain in equilibrium; d) the system starts in thermal equilibrium and is driven from that equilibrium by an external perturbation. Assumption a) and b) are satisfied in many systems described by \eqref{physics}. Assumption c) abd d) may be satisfied e.g. if the system is in contact with one heat bath at temperature $ T $. In particular, assumption d) is satisfied e.g. if the external world raises slowly $ \varepsilon $ starting from a zero value at $ t \rightarrow - \infty $. However, these assumptions are rather artificial, and their validity in many applications is questionable, to say the least. Accordingly, our argument is just a qualitative support, rather than a rigorous proof, of \eqref{variationaloscillating}. 

In conclusion, we discuss a result which seems to agree - qualitatively at least - with the validity of \eqref{variationaloscillating} for stable relaxed states of a system which describes cyclic population dynamics of $ N = 3 $ species, which undergoes a Markovian stochastic process and which is described by an approximate Fokker-Planck equation \cite{Andrae}. The space of the concentrations of 3 chemical species is our $ \left( N = 3 \right)-$state space. It turns out that when the external world changes a control parameter and drives the system across a supercritical Hopf bifurcation from large limit-cycle-like oscillations to small erratic oscillations around a fixed point, then $ \dot{S} $ peaks very much near the transition, and decreases away from it. In our language, it is the amplitude of the oscillation which is in steady state in both configurations; the limit cycle and the fixed point are equivalent to a stable oscillation with constant amplitude $ \neq 0 $ and $ = 0 $ respectively, once small-scale, fast fluctuations have been averaged out. The words 'approximate Fokker-Planck' are justified because the diffusion coefficient is not exactly constant; however, once the system has relaxed then fluctuations of diffusion coefficient too can be averaged out. No exchange with the external world is considered, then $ \Phi = 0 $ and \eqref{bilanciodientropia} gives $\dot{S} = \Pi$. As the system is driven from one stable configuration to the other one, firstly the entropy production increases, then decreases: this is consistent with our result that each stable state corresponds to a local minimum of entropy production. Admittedly, this is no rigorous result, as the value of $ \Pi $ computed when fluctuations are taken into account differs from the value of $ \Pi $ computed when fluctuations are neglected (but in particular cases). Remarkably, however, fluctuations may even \emph{lower} $ \Pi $ \cite{Wachtel}, so that our arguments are not weakened.

\section{More about the pivotal system}
\label{CGLEapp}

A pivotal system satisfies \eqref{slowlyvaryingf}, \eqref{physicssimplified} and, as a consequence, \eqref{C} (or, equivalently, \eqref{London} if $ \mu \neq 0 $). In order to fix the ideas, we start with the DPF filamentary structure discussed in Sec. \ref{SEC5} . At the end, we shall see that a suitable rescaling maps this structure onto somewhat completely different. To start with, let us introduce a complex-valued function 
$ \psi \equiv \sqrt{f} \cdot \exp \left( i \frac{\chi}{\gamma}\right) $ of position where $ \gamma $ 
is $ \propto $ the ('Bohm') diffusion coefficient 
$ D = T \cdot \left(16 \mu \vert \textbf{B} \vert\right)^{-1} $ \cite{Taylor}. We neglect $ \nabla \gamma $; this assumptions is justified below. 
We invoke the following two results of Ref. \cite{Froehlich}. 
Firstly, Ampère's law ensures that both \eqref{London} and $ \textbf{j} = \mu \textbf{J} $ are satisfied provided that 
$ \textbf{J} = \dfrac{\gamma}{2i} \left( \psi^{*} \nabla \psi - \psi\nabla\psi^{*} \right) 
- \mu f \textbf{A} $. 
Secondly, both \eqref{physicssimplified} and the momentum balance (including Lorenz force density) are satisfied, provided that: a) a contribution $ \propto \Delta f $ to the pressure of the fluid (which appears in equation (5) of \cite{Froehlich} and Sec. 8 of \cite{Di Vita}) is neglected, in agreement with \eqref{slowlyvaryingf}; b) viscous terms in the momentum balance are of the same order of the interaction with external currents in Sec. \ref{SEC5} and are neglected, in agreement with equations (3) of \cite{Froehlich} and (8.3) of \cite{Di Vita}; c) $ \psi $ solves the PDE problem:

\begin{equation}
\label{GL}
\begin{cases}
\dfrac{\gamma}{i} \dfrac{\partial \psi}{\partial t}  
-\dfrac{\gamma^2}{2} \left(\nabla - i\dfrac{\mu}{\gamma} \textbf{A}\right)^2 \psi
- \dfrac{1}{\kappa f_{0}} \psi 
+ \dfrac{1}{\kappa f_{0}^2} \vert \psi \vert^2 \psi
- \dfrac{\beta}{f_{0}} \psi \nabla^2 \left( \vert \psi \vert^2 \right)
= 0
\quad ; \\
\\
\textbf{n} \cdot \left(\nabla - i\dfrac{\mu}{\gamma} \textbf{A}\right) \psi = 0
\quad
\mbox{at the boundary}
\end{cases}
\end{equation}

where $ c_{s} $ is the speed of sound and 
$ \kappa \equiv f_{0}^{-1} \cdot c_{s}^{-2}$, $ f_{0} $ is the value of $ f $ in the limit 
$ \textbf{A} \rightarrow 0 $, 
$ \gamma \rightarrow 0 $, $ \beta \rightarrow 0 $. A non-zero value for the quantity $ \beta $ (which has the dimension of $ \gamma^2 $) takes into account possible dissipation-free dispersion of waves across the system; it leaves therefore our discussion concerning  $ \Pi $ unaffected. As for the equation in \eqref{GL}, it is just equation (9) of \cite{Froehlich}, and contains no viscous term. As for the boundary condition, we refer to Sec. 45 of \cite{Landau} which deals with a $ \beta = 0 $ case and which corresponds precisely to vanishing flow across the boundary; this makes sense as \eqref{GL} leads to both \eqref{London} and to the balance of momentum regardless of the actual value of $ \beta $, thus allowing a complete description of the pivotal system. 

In particular, if $ \beta = 0$ - like e.g. in \cite{Di Vita} - then according to \cite{Froehlich} the equation in \eqref{GL} is formally similar to the Ginzburg-Landau ('GL') equation for a superconductor \cite{Landau}. Remarkably, dissipation in a superconductor occurs mainly near the axis of a filament, i.e. where $ \vert \nabla_{y} \ln \left( f \right) \vert \gg 1$. In contrast, the opposite, 'far-from-the-filament-axis' limit \eqref{slowlyvaryingf} corresponds to weak dissipation, in agreement with \eqref{slowdiffusion} as anticipated. (Even if $ \mu \neq 0 $ the main entropy-raising mechanisms in the bulk of of the pivotal system, namely Joule heating and viscous heating, are related to diffusion of magnetic field lines and to diffusion of particles respectively. Both processes lead to dispersion of matter across space. Then, it is self-consistent to neglect dissipation-related terms in the equations of motion when \eqref{slowdiffusion} holds). As shown in Sec. \ref{SEC4}, this does not prevent $ \Pi $ from being $ > 0 $. In the same limit, terms 
$\propto O \left( \vert \nabla \gamma \vert \right) \propto O \left( \vert \nabla \textbf{B} \vert \right)$ become small - as anticipated above - and $ \textbf{A} $ is ruled by the contribution of the external world, which is negligible in the pivotal system. (Gradients of $ \beta $, $ \kappa $ and $ \mu $ vanish everywhere). Finally, if $ \mu \rightarrow 0$, then 
$ \textbf{A} $ plays no role and we retrieve an unmagnetized fluid where Euler equation plays the role of momentum balance.

Now, let us further discuss \eqref{GL} in the 'far-from-filament-axis' region, i.e. in the slow diffusion approximation. To this purpose, we need some intermediate relationships. Firstly, we introduce the quantities 
$ \textbf{c} \equiv \mu \sqrt{\kappa f_{0}} \textbf{A}$,
$ r \equiv \frac{\gamma \sqrt{\kappa f_{0}}}{L}$, 
$ s \equiv \nabla_{y} \cdot \textbf{c} $, 
$ t^{*} \equiv \frac{t}{\gamma \kappa f_{0}} $ and 
$ \eta \equiv \frac{\psi}{\sqrt{f_{0}}} $ (so that 
$ \vert \psi \vert ^2 = f_{0} \vert \eta \vert ^2$). 
Secondly, the definition of $ \psi $ and \eqref{C} imply $f = \vert \psi \vert^2$ and: 

\begin{equation} \label{gradpsi}
\nabla \psi = \left[ \frac{\nabla \ln \left( f \right)}{2} 
+ i \frac{\nabla \chi}{\gamma} \right] \psi
\end{equation}

\begin{equation} \label{gradPhi}
\nabla \chi = \frac{\gamma}{2 i f} \left( \psi^{*} \nabla \psi - \psi\nabla\psi^{*} \right)
= f^{-1} \textbf{J} + \mu \textbf{A}
\end{equation}

The 3 equations \eqref{physicssimplified}, \eqref{gradpsi} and \eqref{gradPhi} lead to the 3 following relationships: 

\begin{equation} 
\label{irc} 
i r \left( \textbf{c} \cdot \nabla_{y} \right) \psi = 
\dfrac{i r}{2} \left[ \textbf{c} \cdot \nabla_{y} \ln \left( f \right) \right] \psi
- \mu \kappa \left( f_{0} f^{-1} \right) 
\left[ \textbf{A} \cdot \textbf{J} + 
\mu \vert \textbf{A} \vert ^2 \vert \psi \vert^2 \right] \psi
\end{equation}

\begin{equation} \label{laplquadr}
\left( f_{0} f^{-1} \right)
\dfrac{\beta}{f_{0}} \psi \nabla^2 \left( \vert \psi \vert^2 \right)
= 
2 \beta \nabla^2 \psi 
- \dfrac{2 i \beta}{\gamma} \psi \nabla^2 \chi
+ \dfrac{\beta \psi}{2} \vert \nabla \ln \left( f \right) \vert ^2
+ \dfrac{2 \beta}{\gamma^2} \psi \vert \nabla \chi \vert ^2
- \dfrac{2 i \beta}{\gamma} \psi \nabla \chi \cdot \nabla \ln \left( f \right)
\end{equation}

\begin{equation} \label{dueibeta}
- \dfrac{2 i \beta}{\gamma} \psi \nabla^2 \chi
= 
\dfrac{2 i \beta}{\gamma f} \psi \dfrac{\partial f}{\partial t}
+ \dfrac{2 i \beta}{\gamma f} \psi \textbf{J} \cdot \nabla \ln \left( f \right)
- \dfrac{2 i \beta}{\gamma} \psi \mu \left( \nabla \cdot \textbf{A} \right)
\end{equation}

where \eqref{slowdiffusion} and \eqref{slowlyvaryingf} imply $ \left( f_{0} f^{-1} \right) \approx 1 $ and $ \vert \frac{1}{f} \frac{\partial f}{\partial t} \vert = O \left( \frac{1}{\tau} \right) \approx 0 $ respectively. Equations \eqref{GL}, \eqref{gradPhi}, \eqref{irc}, \eqref{laplquadr} and \eqref{dueibeta} lead to: 

\begin{eqnarray} \label{GL44}
\begin{split}
&
i \dfrac{\partial \eta}{\partial t^{*}} 
= 
\dfrac{r^2}{2} \left( 1 + \dfrac{\beta}{4 \gamma ^2} \right) \nabla_{y} ^2 \eta +
\left(1 - \vert \textbf{c} \vert ^2 
\right)
\vert \eta \vert ^2 \eta 
+ 
i r 
\left[
\left( 
s + 
\textbf{c} \cdot \nabla _{y} \ln \left( f \right)
\right) 
\left( 
\dfrac{1}{2} + \dfrac{2 \beta}{\gamma ^2} 
\right)
\right] 
\eta 
+
\\
&
+ 
\left[ 
\dfrac{\vert \textbf{c} \vert ^2}{2} - 1 - \kappa \mu \textbf{A} \cdot \textbf{J} -
\dfrac{\beta r^2}{2 \gamma ^2} \vert \nabla_{y} \ln \left( f \right) \vert ^2 - 
\dfrac{2 \beta r^2}{\gamma ^2} \vert \nabla_{y} \left(\dfrac{\chi}{\gamma}\right) \vert ^2
\right] \eta
\\
\end{split}
\end{eqnarray}

Now, we take $ \nabla \cdot \textbf{A} = 0 $ (Coulomb gauge), hence $ s = 0 $. We neglect also $ \vert \nabla_{y} \ln \left( f \right) \vert $ because of \eqref{slowlyvaryingf}. Finally, we note that $ r \propto \gamma \propto D $ and that $ \kappa \propto c_{s}^{-2} \propto T^{-1} \propto D^{-1} $. Then, \eqref{slowdiffusion} allows us to write 
$ \vert \kappa \mu \textbf{A} \cdot \textbf{J} \vert \gg \vert 
\dfrac{2 \beta r^2}{\gamma ^2} \vert \nabla_{y} \left(\dfrac{\chi}{\gamma}\right) \vert$, and \eqref{GL44} reduces to:  

\begin{equation} \label{GL4444}
i \dfrac{\partial \eta}{\partial t^{*}} 
= 
\dfrac{r^2}{2} \left( 1 + \dfrac{\beta}{4 \gamma ^2} \right) \nabla_{y} ^2 \eta +
\left(1 - \vert \textbf{c} \vert ^2 
\right)
\vert \eta \vert ^2 \eta 
+ 
\left[ 
\dfrac{\vert \textbf{c} \vert ^2}{2} - 1 - \kappa \mu \textbf{A} \cdot \textbf{J} 
\right] 
\eta
\end{equation}

Equation \eqref{GL4444} is equivalent to a reduced form of the equation in \eqref{GL} in the slow diffusion approximation. It enjoys two relevant properties. Firstly, in a pivotal system the fact that $ \Pi $ depends explicitly on neither $ \textbf{A} $, $ \beta $, $ \gamma $ and $ c_{s} $ allows the latter quantities to take arbitrary values independently from each other, and \eqref{slowlyvaryingf} allows us to neglect their gradients across the system; by doing so, we make an error 
$ \approx O \left( \vert \nabla_{y} \ln \left( \vert \eta \vert \right) \vert \right) \ll 1$. Secondly, if $ \mu \rightarrow 0 $ then $ \textbf{c} \rightarrow 0$ and the scaling $ \gamma \propto D \cdot \vert \mu \vert^{-1} $ gives $ \gamma \rightarrow \infty $, $ \dfrac{\beta}{\gamma^2} \propto \gamma^{-2} \rightarrow 0$ and $ r \propto \gamma \rightarrow \infty $. Then, \eqref{GL4444} reduces to a particular case of nonlinear Schroedinger equation, i.e. $ -\frac{r^2}{2} \nabla_{y}^2 \eta + \vert \eta \vert ^2 \eta = 0$. Not surprisingly, equations (6), (53) and (54) of \cite{Aronson} show that this equation leads precisely to the Euler equation for unmagnetized fluids, as a term $ \propto \nabla_{y} \ln \left( f \right)$ (the so called 'quantum pressure') is discarded because of \eqref{slowlyvaryingf}. 

At last, here it comes the fundamental result of this Section. Let us take $ \vert \textbf{c} \vert > 1 $ with no loss of generality. Then, in steady state ($ \frac{\partial \eta}{\partial t^{*}} = 0 $ ) \eqref{GL4444} is the steady state version of a GL equation with real coefficients. Now, the latter equation, which we write in the form 

\begin{equation} \label{GL555}
A \zeta + B \nabla_{q}^2 \zeta - C \vert \zeta \vert ^2 \zeta = 0
\end{equation}

describes also the steady state of a fluid near the onset $ Ra = Ra_{cr} $ of Rayleigh-Bénard convection - see equation (5) of \cite{Aronson} for the case $ A = B = C = 1$ and equation (3) of \cite{Wesfried} for the general case respectively; here $ A, B $ and $ C $ are positive constants, $ \zeta \propto \eta $ and $ \nabla_{q}^2 $ is the Laplacian on some rescaled spatial coordinate $ \textbf{q} $. Depending on the values of $ \mu $ and $ \gamma $, the boundary condition on $ \zeta $ may be either Dirichlet ($ \frac{\mu}{\gamma} \ll 1 $), Neumann ($ \frac{\mu}{\gamma} \gg 1 $) or a linear combination of the two ($ \frac{\mu}{\gamma} \approx 1 $). Together, the fact that $ \eta \propto \psi \propto \sqrt{f} $ and equation \eqref{densitascalare} allow us to find a rescaling which maps \eqref{GL4444} in steady state and \eqref{GL555} onto each other (gradients of $ \vert \eta \vert $ are negligible in the slow diffusion approximation, because of  \eqref{slowlyvaryingf}). In particular, once both $ r $, $ \frac{\beta}{\gamma^2} $, $ \textbf{c} $ and $ \kappa \mu \textbf{A} \cdot \textbf{J}$ are known then \eqref{GL4444} in steady state reduces to \eqref{GL555} provided that we identify $ A, B, C, \zeta $ and $ \textbf{q} $ with 
$ \frac{\vert \textbf{c} \vert ^2}{2} - 1 - \kappa \mu \textbf{A} \cdot \textbf{J} $, 
$ \frac{r^2}{2} \left( 1 + \frac{\beta}{4 \gamma ^2} \right) $,
$ \vert \textbf{c} \vert ^2 - 1 $, 
$ \eta $ and $ \textbf{y} $ respectively.

The reverse is also true. All the way around, once $ A, B $ and $ C $ are known for a fluid described by the quantity $ \zeta = \zeta \left( \textbf{q} \right) $ in the $ \textbf{q}-$space near the convection onset and a known boundary condition (Dirichlet, Neumann or a linear combination of the two). Then, it is always possible to map \eqref{GL555} onto a steady state version of \eqref{GL4444} of a pivotal system through a rescaling with Jacobian $ \Lambda $, because we have 5 unknown quantities ($ \Lambda $, $ r $, $ \frac{\beta}{\gamma^2} $, $ \textbf{c} $ and $ \kappa \mu \textbf{A} \cdot \textbf{J}$) against the 4 = 3 + 1 constraints given by the 3 known values of $ A, B $ and $ C $ and by the 1 boundary condition (which gives the value of $ \frac{\mu}{\gamma} $). There is still room for 5 - 4 = 1 degree of freedom, namely $ \Lambda $.

In other words, a steady state fluid on the verge of Rayleigh-Bénard convection is in the same class of equivalence of a particular pivotal system, namely the steady, relaxed filamentary configuration of a DPF pinch. It is precisely the fact that dissipation is not required for the description of the steady relaxed state - see Sec. \ref{SEC4} - which allows the inviscid equation \eqref{GL} to provide information on the steady relaxed configurations of fluids with convection. Rather, dissipation describes relaxation; the latter is very slow in the slow diffusion approximation and may therefore safely be described as a slow succession of steady states. 

Of course, detailed dynamics is completely different in the two systems: DPF filaments are 'steady' only when compared with MHD time-decay, the temperature gradient across the fluid is not negligible, radiation transport of heat occurs across the DPF, and is it is not even required that a magnetic field occurs in the Rayleigh-Bénard case. But dynamics provides just constraints on minimization of $ \Pi $; it does not change the fact that if a steady state is stable in one system - hence $ \Pi = \min $ - then the corresponding steady state is also stable in the other system - i.e., $ \Pi $ remains a minimum. This is in analogy to what happens at thermodynamic equilibrium, where the fact that $ S $ attains a maximum does not depend on the choice of the $ \textbf{x} $'s. We may say that the problems of stability of the electric current filaments in a DPF pinch and of a fluid near Rayleigh-Bènard convection are dual to each other.

\section{More about \eqref{Rossby}}
\label{RossbyAPP}

Independent estimates of $ \emph{x} $ confirm \eqref{Rossby}. The plasma of our shell is made of has three remarkable features \cite{Priest}: a) inter-particle collsions are far from negligible, as $ \eta_{*} > 0 $; b) the pressure is not negligible in comparison with the magnetic energy density $ \frac{\vert \textbf{B} \vert^2}{2 \mu_{0}} $; c) radiative transport of heat is relatively weak. In such plasmas, a rescaling $ \textbf{x} \rightarrow \xi \textbf{x} $ by a dimensionless factor $ \xi $ leaves physics unaffected provided that the Collisional Vlasov High Beta Scaling ('CVHBS') holds \cite{ConnorTaylor}: 

\begin{equation}
\label{CVHBS}
\textbf{x} \rightarrow \xi \textbf{x}
\qquad 
;
\qquad 
t \rightarrow \xi^{\dfrac{5}{4}} t
\qquad 
;
\qquad 
\textbf{B} \rightarrow \xi^{-\dfrac{5}{4}} \textbf{B}
\end{equation}

According to \eqref{CVHBS}, $ D \rightarrow \xi^{\dfrac{3}{4}} D $ and 
$ \tau \rightarrow \xi^{\dfrac{5}{4}} \tau $, 
as $ D $ and $ \tau $ have the dimensions of (length)$^{2}/$(time) and of time  respectively. Together, \eqref{Rossby} and \eqref{CVHBS} imply:

\begin{equation}
\label{tauRo}
\tau \propto Ro
\end{equation}

i.e., the relaxation time of a perturbation of our shell increases with increasing Rossby number. The results of \cite{Bercik}, which do not rely on our assumption 
$ \Delta R \ll \frac{T}{\vert \nabla T \vert} $, confirm \eqref{tauRo}.

We have neglected viscosity heating so far. Numerical simulations of an extensive set of dynamo models in rotating spherical shells \cite{Christensen} take into account also viscous heating and cover a wide range of control parameters. Generally speaking, these simulations hint at lack of explicit dependence of the magnetic field on either resistivity or rotation period, and suggest rather a strong dependence on buoyancy, in agreement with our discussion. Results are fitted by the following scaling laws:

\begin{equation}
\label{Christensen}
Ro_{*} \propto Ra_{*}^{\alpha_{R}}
\qquad 
;
\qquad
Nu_{*} \propto Ra_{*}^{\alpha_{N}}
\qquad 
;
\qquad
\alpha_{R} \approx \dfrac{2}{5}
\qquad 
;
\qquad
\alpha_{N} \approx \dfrac{1}{2}
\end{equation}

where 
$ Ro_{*} \equiv \frac{u_{c} \emph{P}}{\Delta R} = \frac{L_{c} Ro}{\Delta R} $, 
$ Nu_{*} \equiv \frac{Q_{ad} \emph{P}}{\Delta R^{3} 
\left[ T \left( R_{\min} \right) - T \left( R_{\max} \right) \right]} $ and 
$ Ra_{*} \equiv \frac{Q_{ad} \emph{P}^3}{\left( \Delta R \right) ^{4}} $ are a modified Rossby number, a modified Nusselt number and a modified Rayleigh number representing the cell rotation, the heat transport and the (suitably normalized) amount $ Q_{ad} $ of heat advected by buoyancy per unit volume and unit time respectively. 
Together, \eqref{Christensen} and the definitions of $ Nu_{*} $ and $ Ra_{*} $ give:

\begin{equation}
\label{Qad}
Q_{ad} \propto
\emph{P}^{\left( \dfrac{3 \alpha_{N} - 1}{1 - \alpha_{N}} \right)}
\left[ T \left( R_{\min} \right) - T \left( R_{\max} \right) \right]^{\left( \dfrac{1}{1 - \alpha_{N}} \right)}
\left( \Delta R \right) ^{\left( \dfrac{3 - 4 \alpha_{N}}{1 - \alpha_{N}} \right)}
\end{equation}

Now, \eqref{CVHBS} imply

\begin{equation}
\label{Tscaling}
T \rightarrow \xi^{-\dfrac{1}{2}} T
\end{equation}
 
because $ T $ has the dimensions of energy 
$ \propto $ (length)$^{2}/$(time)$^{2}$ 
(or,equivalently, of $ \vert \textbf{B} \vert D $). Relationships \eqref{CVHBS}, \eqref{Qad} and \eqref{Tscaling} give:

\begin{equation}
\label{Qscaling}
Q_{ad} \rightarrow \xi^{\left( \dfrac{5 - \alpha_{N}}{4 - 4 \alpha_{N}} \right)} Q_{ad}
\end{equation}
 
Together, \eqref{CVHBS}, \eqref{Christensen}, \eqref{Qscaling} and the definition of $ Ra_{*} $ give:

\begin{equation}
\label{Rascaling}
Ra_{*} \rightarrow \xi^{\left( \dfrac{5 - \alpha_{N}}{4 - 4 \alpha_{N}} - \dfrac{1}{4}\right)} Ra_{*}
\end{equation}

Moreover, we assume that a small mass element follows a random walk among many convective cells and write $ \frac{Ro_{*}}{Ro} = \frac{L_{c}}{\Delta R} \propto \sqrt{t} $, so that \eqref{CVHBS} implies:

\begin{equation}
\label{Brownian}
\dfrac{Ro_{*}}{Ro} \rightarrow \xi^{- \dfrac{5}{8}} \dfrac{Ro_{*}}{Ro}
\end{equation}

Finally, \eqref{Rossby}, \eqref{CVHBS}, \eqref{Christensen}, \eqref{Brownian} and \eqref{Rascaling} and the definition of $ Ro_{*} $ give 
$ \emph{x} = \dfrac{10 \left(1 - \alpha_{N} \right)}{5 \left(1- \alpha_{N} \right) + 8 \alpha_{R}} \approx 0.88 $, a result which is still compatible with the range $ 0.78 < \emph{x} < 1.9 $ reported in \cite{Jordan000}. 

\end{document}